\documentclass[aps,prl,nobibnotes,twocolumn,showkeys,superscriptaddress]{revtex4-1}

\usepackage[artemisia]{textgreek}
\usepackage{amssymb}
\usepackage{epsfig}
\usepackage{amsmath}
\usepackage{graphicx}
\usepackage{dcolumn}
\usepackage{latexsym}
\usepackage{color}
\usepackage{epstopdf}
\usepackage{subfigure}
\usepackage{nicefrac}
\usepackage{blindtext}
\usepackage[ulem=normalem]{changes}
\usepackage{float}
\usepackage{xcolor}
\usepackage{soul}
\usepackage[T1]{fontenc}
\usepackage[hidelinks]{hyperref}
\usepackage{xcolor}
\setcitestyle{super}
\hypersetup{
    colorlinks,
    linkcolor={red!50!black},
    citecolor={blue!50!black},
    urlcolor={blue!80!black}
}
\usepackage{lipsum}
\usepackage[utf8]{inputenc}

\usepackage[symbol]{footmisc}

\begin{document}


\title{Phase controlled multi-terminal Josephson junction in ternary hybrid nanowire}






\author{Sabbir A. Khan$^\ddag$}
\email{skh@dfm.dk}
\altaffiliation{$^\ddag$Contributed equally to this work}
\affiliation{Center for Quantum Devices, Niels Bohr Institute, University of Copenhagen, 2100 Copenhagen, Denmark}
\affiliation{Danish Fundamental Metrology, Kogle Alle 5, 2970 Hørsholm, Denmark}

\author{Lukas Stampfer$^\ddag$}
\affiliation{Center for Quantum Devices, Niels Bohr Institute, University of Copenhagen, 2100 Copenhagen, Denmark}

\author{Sara Mart\'{i}-S\'{a}nchez}
\affiliation{Catalan Institute of Nanoscience and Nanotechnology (ICN2), CSIC and BIST, Campus UAB, Bellaterra, 08193 Barcelona, Catalonia, Spain}

\author{Dags Olsteins}
\affiliation{Department of Energy Conversion and Storage, Technical University of Denmark, Fysikvej, Building, Lyngby, 310, 2800 Denmark}

\author{Damon James Carrad}
\affiliation{Department of Energy Conversion and Storage, Technical University of Denmark, Fysikvej, Building, Lyngby, 310, 2800 Denmark}

\author{Thies Jansen}
\affiliation{Department of Energy Conversion and Storage, Technical University of Denmark, Fysikvej, Building, Lyngby, 310, 2800 Denmark}

\author{Jonas Johansson}
\affiliation{Solid State Physics and NanoLund, Lund University, P O Box 118, SE-221 00 Lund, Sweden.}



\author{Jordi Arbiol}
\affiliation{Catalan Institute of Nanoscience and Nanotechnology (ICN2), CSIC and BIST, Campus UAB, Bellaterra, 08193 Barcelona, Catalonia, Spain}
\affiliation{ICREA, Pg. Lluís Companys 23, 08010 Barcelona, Catalonia, Spain}

\author{Peter Krogstrup}
\affiliation{NNF Quantum Computing Programme, Niels Bohr Institute, University of Copenhagen, Denmark}

\author{Thomas S. Jespersen}
\email{tsand@nbi.ku.dk}
\affiliation{Center for Quantum Devices, Niels Bohr Institute, University of Copenhagen, 2100 Copenhagen, Denmark}
\affiliation{Department of Energy Conversion and Storage, Technical University of Denmark, Fysikvej, Building, Lyngby, 310, 2800 Denmark}

\date{\today}

\begin{abstract}

This work presents multiterminal Josephson junctions in hybrid semiconductor-superconductor InAsSb-Al nanocrosses. Hybrid nanocrosses are grown using molecular beam epitaxy and are formed through As-assisted merging of oppositely directed InAsSb nanowires. We explain this complex ternary merging mechanism using a temperature-dependent phase diagram and investigate the detailed crystal structure with atomic-resolution imaging. The hybrid nanoscrosses enabled the fabrication of multiterminal Josephson junction devices, which were characterized at low temperatures. The supercurrent through each terminal combination was measured as a function of the density in the junction and the relative phase of the terminals, which was controlled by an external magnetic field.

\end{abstract}


\maketitle

Expanding the number of terminals (N>2) in the conventional Josephson Junction (JJ) by connecting multiple superconductor leads through a common non-superconducting region small enough to allow formation of bound states between all electrodes. Such Multi-terminal JJ (MJJ) \cite{riwar2016multi} configuration enhances control over quantum states and facilitates advanced gate operations for superconducting circuits. Furthermore, the multiterminal platform holds the potential for quasiparticle braiding, towards robust and fault-tolerant qubits \cite{pankratova2020multiterminal}. Lately, several materials platforms have been considered to explore MJJ, including graphene \cite{draelos2019supercurrent, zhang2023andreev} and planar semiconductor/insulator-superconductor heterostructures \cite{schiela2024progress, zhang2024large}. In the planar platform, top-down processing is required to create MJJ, which allows freedom and scalability; however, it also comes with constraints due to the processing-related damages in the junctions \cite{khan2020highly, carrad2020shadow, paudel2025disorder}. In addition to the planar platform, 1D bottom-up \cite{khan2021multiterminal, rossi2024stemless} and selective area-grown branched networks have recently been experimented with for the multiterminal JJ platform \cite{KrizekSAG2018, friedl2018template, aseev2019ballistic}. Although limited by the flexibility of MJJ design and scalability, the 1D network provides intrinsic materials purity for fundamental study \cite{khan2021multiterminal}.


\begin{figure*}[ht!]
\includegraphics[width=0.93\textwidth]{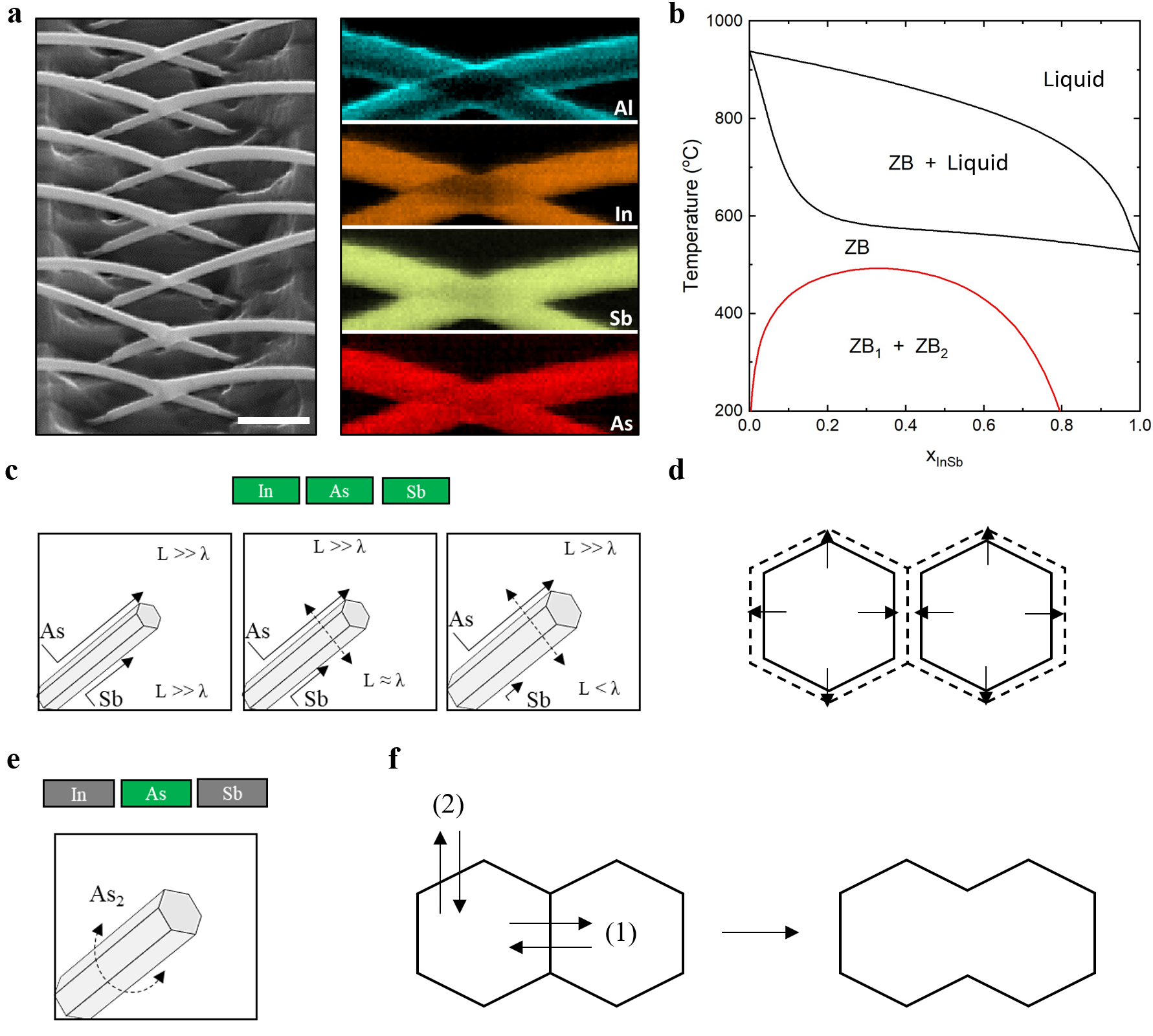}
\caption{{\textbf{Formation of the ternary network.}} \textbf{a}, Scanning electron microscopy of InAsSb-Al nanocrosses grown from the V-grove trenches. To the right, Electron Energy Loss Spectroscopy elemental maps show the spatial distribution of In, As, Sb and Al. \textbf{b}, The Phase diagram of the possible crystal structure of ternary NW as a function of Sb composition and growth temperature. With growth temperature around 450$^{\circ}$ C and intended Sb percentage of 60-70\% we expected to achieve ZB crystal. \textbf{c}, Schematic of the NW growth in different stages of As and Sb diffusion length. All the fluxes are open and As assists in axial growth whereas later stage Sb contributes radial broadening. \textbf{d}, Sb driven radial growth helps NWs to get closer and possibly connect in through side-wall. \textbf{e}, Post growth radial broadening with additional As flux to merge the NWs properly. \textbf{f}, Schematic of two As-assisted integration mechanisms (referred as (1) and (2)) simultaneously occur to merge the NW.  Scale bars are: (a) 1 \textmu m.}
\label{fig1}
\end{figure*}



Different 1D materials have recently been explored, including InAs and InSb \cite{liu2019semiconductor, khan2020highly, zhang2022signatures, khan2023epitaxially, carrad2022photon, chen2023gate, stavenga2023lower, goswami2023sn, badawy2023tunable}. Although the InAs-based 1D network offers the advantage of robust device fabrication, it also induces a crystal phase change at the junction, thereby disrupting coherence. Conversely, the InSb-based 1D network maintains high crystal quality with a single phase but is limited by intrinsic constraints such as the limited diffusion length, resulting in constraints on the size flexibility of the 1D network. Furthermore, InSb-based networks are sensitive to high temperatures and chemically fragile, imposing limitations on temperature-dependent operations and processing such as selectively etching of grown superconductors \cite{khan2021multiterminal, rossi2024stemless, plissard2013formation}. Taking these into account and leveraging the strengths of both materials, in this work, we investigated ternary InAsSb-Al nanowire networks for multiterminal Josephson junctions (JJs), which exhibit robustness similar to InAs-Al while maintaining high crystal quality in the junction area like InSb-Al. We have demonstrated the growth of hybrid multiterminal InAsSb-Al using molecular beam epitaxy and provided an analysis of the As-assisted merging for multiterminal junction formation. We also performed atomic-scale analysis and revealed the crystal direction and atomic arrangement throughout the intersection. Next, we used the grown structure to fabricate multiterminal JJ and measured behavior at mK temperatures. We characterized the supercurrent flow in the junction by means of DC transport and demonstrated that each terminal contributes to the supercurrent. Finally, we show the gate behavior of the device, which is critical to facilitate tuning to the single channel regime.


\section{Results and Discussion}

\textbf{InAsSb Nanocrosses Formation.} InAsSb were grown from Au catalysts which were precisely deposited on the angled trenches opposite to each other \cite{khan2020highly, khan2021multiterminal, carrad2022photon, chen2023gate}. To assist the Sb-contained growth, the NWs grownth was initiated by with InAs stem and subsequently the material sources were changed into the ternary composition \cite{khan2020highly, khan2023epitaxially, carrad2022photon, stampfer2022andreev}. Fig.\ref{fig1} (a) shows the array of InAsSb nanocrosses (NCs) grown in a single V-groove trench (approximately 3um lateral distance). Post-growth analysis shows that the InAs stems are buried under the overgrown trenches. The catalyst particle pair in the opposite direction is placed with an offset of approximately 120-150 nm, which is close to the diameter of individual NW. Hence, NW pairs initially maintain a gap between them and merge later  \cite{khan2021multiterminal}. After InAsSb, the sample was cooled down and approximately  15 nm Al was grown on the 3 facets of the crosses. Detail of the hybrid growth is presented in the method section. The electron energy-loss spectroscopy (EELS) analysis of four different elements of the NCs is also shown in Fig.\ref{fig1} (a). We can observe a homogeneous distribution of Al, In, Sb and As through the cross except for the central part where the higher sample thickness reduces the e-beam transmission. Thus, a low signal can be seen in the central region. Complementary analysis was performed by energy dispersive x-ray spectroscopy (EDX) to obtain quantitative data on the NC composition. Supporting information S1 presents the spectrum obtained in the central part of the NC and a quantitative profile of one arm passing through the intermediate area. The composition is found to be, In: 50.19 $\%$, Sb: 30.79 and As 19.01 $\%$.

In Fig.\ref{fig1} (b), we show the pseudobinary phase diagram for InAs$_{1-x}$Sb$_x$. Depending on the temperature and the Sb composition, four different phase regions can be seen: liquid at high temperature, solid zinc blende phase (ZB) in coexistence with liquid at intermediate temperatures, ZB phase at lower temperatures, and a miscibility gap at temperatures below a certain critical temperature and for certain compositions. The miscibility gap is an effect where not all compositions of InAs$_{1-x}$Sb$_x$ being thermodynamically stable at temperatures lower than a certain critical temperature, which is 500$^\circ$C for this materials system.

Guided by the schematics in Fig.\ref{fig1} (c-f), we give a qualitative explanation of the As-assisted merging of the nanowires into nanocrosses. The first step in nanocross formation is to grow adjacent Sb-rich nanowires until they touch each other as indicated in Fig.\ref{fig1} (c-d). All the flux pressure was provided for 35 min and the substrate temperature was kept to approx. 447$^\circ$C in this stage. This stage was performed with As$_4$. After this, to properly merge the nanostructure, the temperature is lowered to approx. 340$^\circ$C and the nanostructures are exposed to As$_2$ for 10 min (Fig.\ref{fig1} (e)), which we believe forces an anion exchange \cite{wang2002thermodynamic} driven by an excess of arsenic from the ambient phase (indicated by process (2) in Fig.\ref{fig1} (f)). Experimental investigations have shown that the As-for-Sb exchange can extend to several monolayers in GaAs \cite{kaspi1999compositional,xie1999arsenic}. In addition to this, an atomistic mechanism for segregation and As-for-Sb exchange in the InSb surface has been proposed earlier \cite{anderson2018atomistic}.

This anion exchange makes the surface composition of the nanowires less Sb rich and eventually drives the composition toward, or even into, the miscibility gap. A composition in the miscibility gap is likely to lead to a phase segregation process where the atoms in the surface layer of the nanowires are highly mobile as long as the As$_2$ is on. This increased mobility could then result in an efficient interdiffusion (process (1)) in Fig.\ref{fig1} (f)) and surface energy minimization of the touching wires into well-merged, single-crystalline nanocrosses.


\begin{figure*}[ht!]
\includegraphics[width=0.93\textwidth]{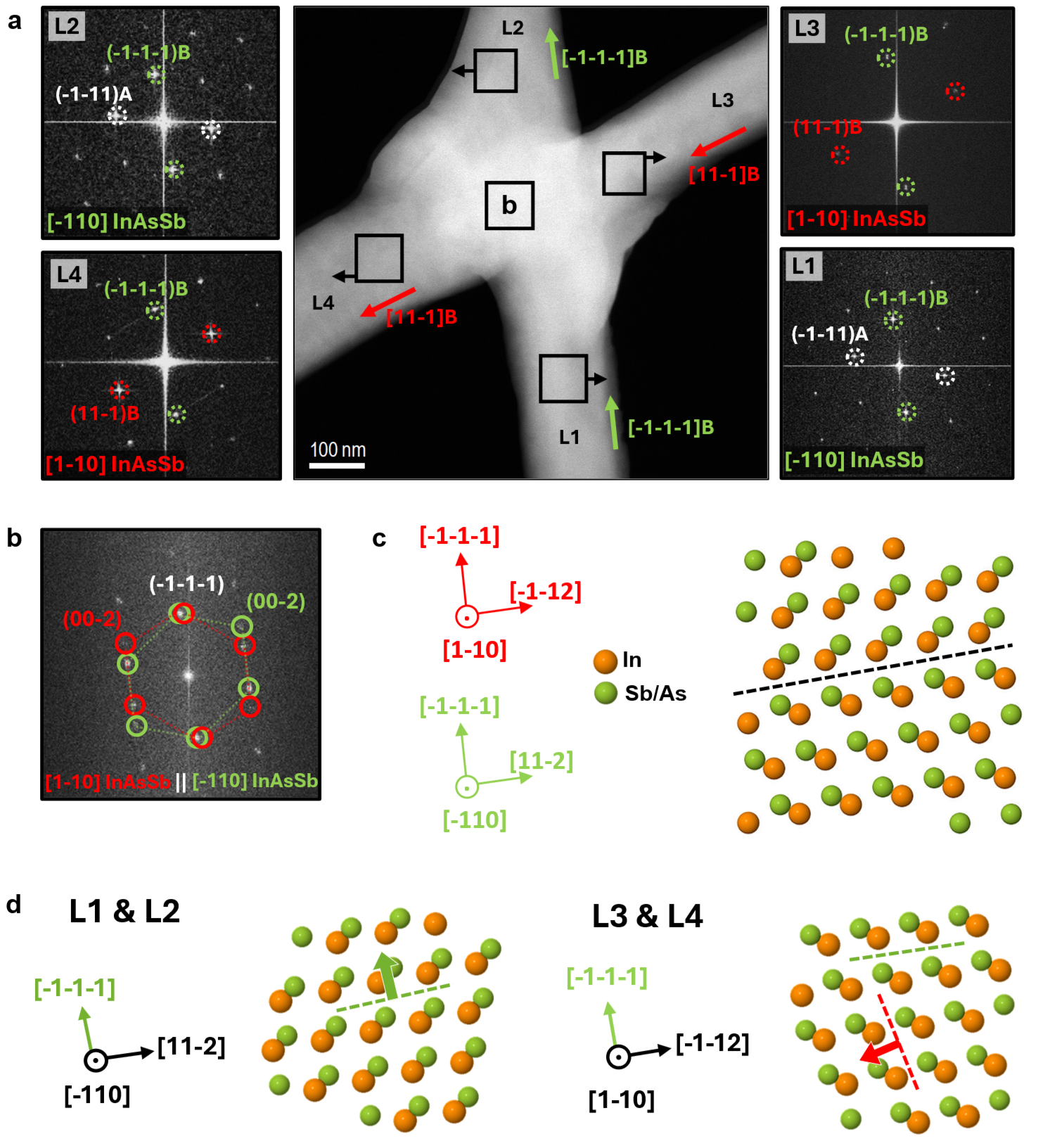}
\caption{{\textbf{Structural analysis of InAsSb nanocrosses.}} \textbf{a}, Low-magnification HAADF-STEM micrograph of the four-terminal InAsSb junction. ‘Green’ and ‘Red’ arrows in the four legs (labelled L1-4) indicate each NW growth direction before and after merging. The power spectrum of micrographs acquired at each terminal are displayed to show the crystal plane directions of every leg. \textbf{b}, FFT from the central area of the junction showing coexistence of both crystal orientations, which join together forming twin boundaries. \textbf{c}, The atomic arrangement across the twin in the central area with the corresponding crystal directions labelled. \textbf{d}, Atomic models showing the crystal orientation in the different terminals of the network.}

\label{fig2}
\end{figure*}


\textbf{Structural Analysis.} Atomic-scale structural analysis of a four-terminal junction is presented in Fig.\ref{fig2} (a). For simplicity, we labeled opposite directional NWs as green (L1 / L2) and red (L3 / L4). By analyzing the FFTs of atomic-resolution images at each one of the legs, we can see how the crystal directions are maintained for each NW through the intersection (L1-L2/L3-L4), although there is a mirror rotation for both red and green NWs. In all branches, there is a common $\{$111$\}$ plane that matches the growth direction of the green wire. On the other hand, the growth plane of the red NW appears in a (111) mirror of the green NW. So, the merging is accompanied by the creation of twin boundaries in the central area, which preserves the continuity of the crystal system by joining both growth directions. In Fig.\ref{fig2} (b), the central area FFT has been indexed showing the twin configuration, and the same behavior can be expanded to the other FFTs obtained at different junction positions. In Fig.\ref{fig2} (c), the atomic arrangement across the twin is shown in the schematic with the corresponding crystal directions. In Supporting Information S2, we provide additional analysis of the junction area, revealing the presence of twins in the (–1–1–1) plane at different edges of junction region. Also, complex defective structures are present at the edge of the junction, where overgrowth is more visible (see Supporting Information S3). In these regions, we observed consecutive twinning in the (-1-11) plane and irregular boundaries corresponding to the merging of differently oriented regions. In addition, the schematic of the atomic arrangement with the corresponding crystal directions of each leg of the cross is shown in Fig.\ref{fig2} (d). 



Finally, in the presence of an MBE-grown Al shell, we found that the outer part of the overgrown region shows a relative compression with respect to the central part of the cross. The compression is measured based on the FFT reflections as -2.25\% to -3.2\%. For further investigation, spatially resolved dilatation analysis with Geometric Phase Analysis (GPA) was performed, but weak transmission as a result of the sample thickness makes fluctuations in the map. Hence, only overall compression with FFT analysis was performed and presented in the Supporting Information S4.


\begin{figure*}
\includegraphics[width=18cm]{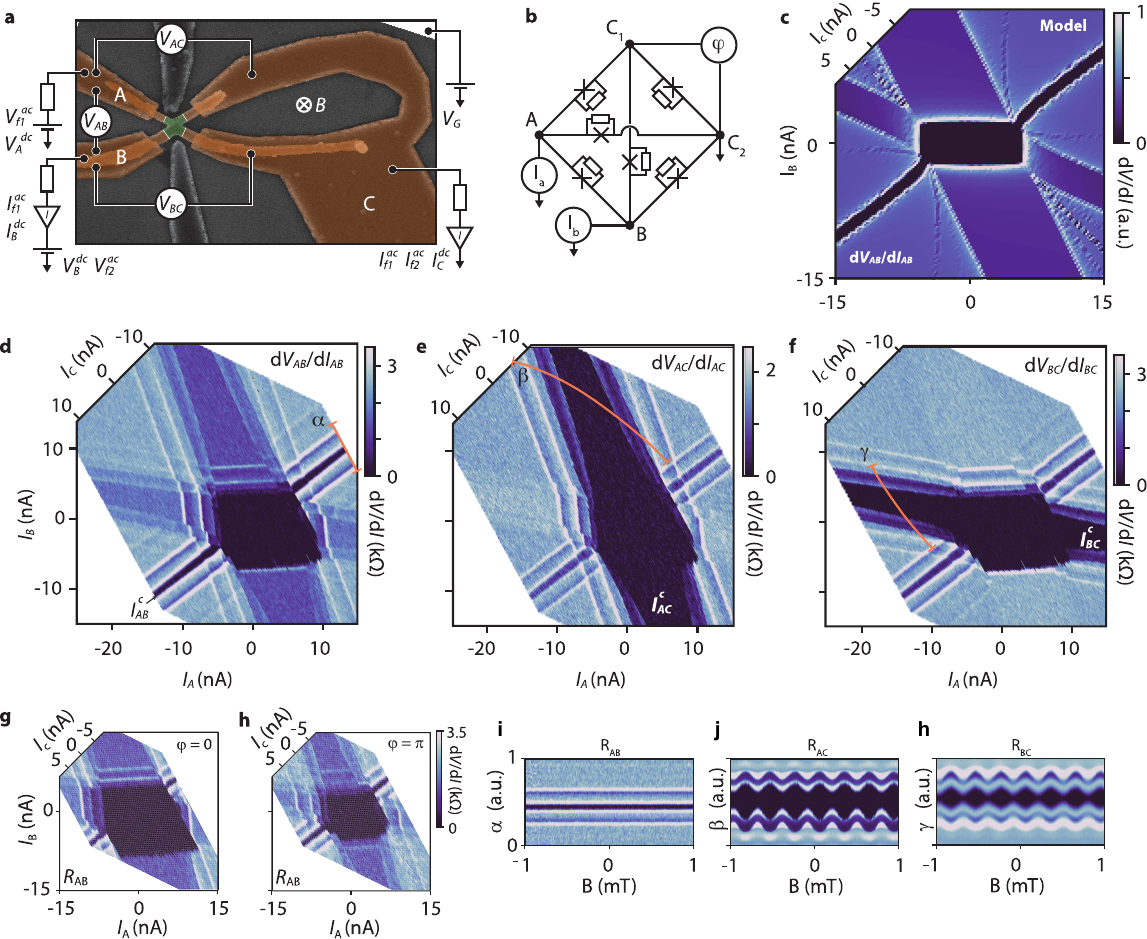}
\vspace{-0.5cm}
\caption{\textbf{Supercurrents in an InAsSb-Al multi-terminal Josephson device.}
\textbf{a}, False colored SEM of a typical device with a schematic of the employed measurement circuit. Superconducting contacts are in orange, and exposed semiconducting InAsSb in green. \textbf{b}, Schematic of RSJ-based circuit used to simulate the overall characteristics of the devices. \textbf{c}, Simulated differential resistance, $R_\mathrm{AB}$, from terminal $A$ to $B$ as a function of the applied currents. \textbf{d-f}, Measured differential resistances from terminal $A$ to $B$ ($R_\mathrm{AB}$), from $A$ to $C$ ($R_\mathrm{AC}$), and from $B$ to $C$ ($R_\mathrm{BC}$), respectively. \textbf{g-h}, Differential resistance, $R_\mathrm{AB}$ measured at two different phase differences $\phi =0$ and $\phi = \pi$, respectively. \textbf{i-h}, The corresponding phase dependence measured along the three paths $\alpha, \beta, \gamma$, as indicated in (d-f). The measurements were performed at $T=20 \, \mathrm{mK}$ and $V_{G}=40$.}
\vspace{-0.5cm}
\label{fig3}
\end{figure*}

\textbf{Multi-terminal Josephson device} 
Al contacted ternary NCs were used to characterize the performance of the NCs as multi-terminal Josephson devices. The Al superconductor was selectively etched in the junction area without damaging the ternary semiconductor. Details of fabrication is presented in the Methods section. Figure \ref{fig3} (a) shows an SEM of a typical device: two terminals, $A,B$, have individual contacts, and the remaining two were connected into a $\sim 5 \, \mu \mathrm{m}^2$ flux pick-up loop, $C$, enabling control of the phase difference using an external magnetic field. Simultaneous measurements of the three differential resistances  $R_\mathrm{AB} \equiv dV_\mathrm{AB}/dI_\mathrm{AB}$, $R_\mathrm{AC} \equiv dV_\mathrm{AC}/dI_\mathrm{AC}$, and $R_\mathrm{BC} \equiv dV_\mathrm{BC}/dI_\mathrm{BC}$ were performed as a function DC currents $I_\mathrm{A},I_{B},I_{C}$ using the dual lock-in measurement setup schematically shown in Fig.\ \ref{fig3}(a). Here $I_\mathrm{B}$ and $I_\mathrm{C}$ were measured directly (and $I_\mathrm{A} = -(I_\mathrm{B} + I_\mathrm{C})$) and tuned by biasing through $R_b = 50 \mathrm k \Omega$ resistors. The phase difference $\phi$ of terminals $C1$ and $C2$ was controlled with an external perpendicular magnetic field $B$ and the overall conductance of the junction was tunable by the potential $V_g$ applied to the doped-Si back-gate. The side gates in Fig.\ \ref{fig3}(a) were not used in this study and kept grounded. Here, we discuss results from one device; however, in total, we measured three devices showing similar behavior, and results from other data can be found in Supporting information S5.

Given superconducting coherence between each pair of terminals, our device can be modeled by the RSJ-circuit shown in figure \ref{fig3}(b), characterized by six critical currents and  corresponding normal resistances. Fig. \ref{fig3}(c) shows the RSJ simulated differential resistance map $R_\mathrm{AB}$ as a function of $I_{A}$ $I_{B}$ and $I_{C}$. The resistance map features a central zero-voltage supercurrent state, where all the resistances $R_\mathrm{AB}$, $R_\mathrm{AC}$ and $R_\mathrm{BC}$ vanish and a zero $R_\mathrm{AB}$ branch, where only $R_\mathrm{AB}$ vanishes when $I_{AB} < I^C_{AB}$. The angles of these branches with respect to the main axes of the map are set by the ratios of the normal-state resistances. In the RSJ-model calculation the values of the normal state resistance and critical currents are based on the experimental measurements to roughly match the experiment.  

The experimental differential resistance maps $R_\mathrm{AB}$, $R_\mathrm{AC}$, and $R_\mathrm{BC}$ of the device are shown in Fig.\ \ref{fig3}(d-f) as a function of $I_{A}$ $I_{B}$ and $I_{C}$. For a given point, the values of $I_A$, $I_B$, and $I_C$ are found by normal projections onto the respective axis. In agreement with the RSJ-model, a zero-voltage supercurrent state appears, and each map shows a distinct zero resistance branch for the corresponding terminals. The critical current for the three branches $I^C_{AB}= 0.5 \, \mathrm{nA}$, $I^C_{AC}= 4.2 \, \mathrm{nA}$, and $I^C_{BC}= 1.8 \, \mathrm{nA}$ are extracted from the width of the branches away from the central zero-resistance region. The critical currents show a weak decrease at larger distances from the zero-current center as previously observed in MJJ and attributed to joule-heating from the dissipative junctions\cite{Draelos:2019}. The normal state resistances are $R^N_{AB} \sim 2.8 \, \mathrm{k} \Omega$, $R^N_{AC} \sim 1.8 \, \mathrm{k}\Omega$, $R^N_{BC} \sim 2.3 \, \mathrm{k}\Omega $ and the product $I_cR_n \sim 5 \mu \mathrm V$ is significantly smaller than ideal theoretical predictions $\Delta_\mathrm{Al}/e \sim 225 \, \mu \mathrm V$ as is also often observed hybrid devices based on single nanowires \cite{khan2020highly}. For the influence of the sweeping direction on the experimental differential resistance maps $R_\mathrm{AB}$, $R_\mathrm{AC}$, and $R_\mathrm{BC}$, see supporting information S6.

The critical currents are larger for the combinations including terminal $C$. We attribute this difference to two effects. Firstly, terminal $C$ consists of two of the 4 cross legs, therefore the total probability for an electron to scatter into one of those arms increases.Secondly, the critical current ”around a corner” is generally smaller compared to the critical current following a straight path, which may be related to crystal phase boundaries that have to be crossed. The two connections to $C$ both have at least one straight junction, while for the $AB$ connection the critical current is carried fully around a crytal-phase boundary. Note that all three branches can be identified in all maps due to voltage divider effects: For example when BC transitions from a finite to a zero-voltage stage the AB differential resistance drops as well. 

While the three supercurrent branches can be considered as conventional 2-terminal JJs, this is not the case for the central region where all three resistances simultaneously vanish. Within this region, the four terminals are at the same voltage, and the multi-terminal junction thus carries supercurrent among all terminals. The shape of the region is set by the intersection of the three individual branches of critical current, and distorted by the voltage-divider effects and  effects of hysteresis where the transition to the zero-voltage state coming from a large current state occurs at the retrapping which is lower than the transition to the finite-voltage state upon increasing the current from zero. 

The $\sim 5~\mu \mathrm m^2$ superconducting loop in Fig.\ \ref{fig3}(a) allows tuning the phase-difference $\phi$ of the two $C$-terminals by a perpendicular magnetic field, $B$. Changes in the field by $\Delta B=\Phi_0/A \sim 320 \, \mu T$ modulate $\phi$ by $2 \pi$. Fig. \ref{fig3} (g,h) shows the current maps of $R_\mathrm{AB}$ for $\phi = 0$ and $\phi = \pi$. Additional current maps for phases in between 0 and $\pi$ are shown in supporting information S7. The size of the central zero voltage region as well as $I^\mathrm{C}_\mathrm{AC}$ and $I^\mathrm{C}_\mathrm{BC}$ of the $AC$- and $BC$-branches decrease upon increasing $\phi$ from 0 to $\pi$. The $AB$-branch, however, is unaffected. This behavior is confirmed further in panels (i-h) showing measurements sweeping along the paths $\alpha, \beta, \gamma$ shown in Fig.\ \ref{fig3}(d-f) while slowly increasing $\phi$. While the $A-B$ critical current stays constant with $\phi$, the modulation of the $A-C$ and $B-C$ critical current resembles that of a three-terminal SQUID-type device \cite{Jaklevic1964Quantum, giazotto2011josephson,savinov2016enhancement,galaktionov2013current}. The observed independence of $I^\mathrm{C}_\mathrm{AB}$ on $\phi$ is also expected since tracing along the $\alpha$ axis, for currents ($I_\mathrm{A}, I_\mathrm{B}, I_\mathrm{C}$) away from the central region, there a dissipative current and thus a non-constant $A-C$ and $B-C$ phase differences and the external flux modulation in the loop does not modulate the $A-B$ branch. Since two terminals of the device share the contact $C$, the individual currents cannot be measured directly. However, the phase modulation of the $A-C$ and $B-C$ critical currents show that $A$ and $B$ terminals are each connected to both $C$-terminals and thus confirm the coherent four-terminal nature of the device. 



\begin{figure}
\includegraphics{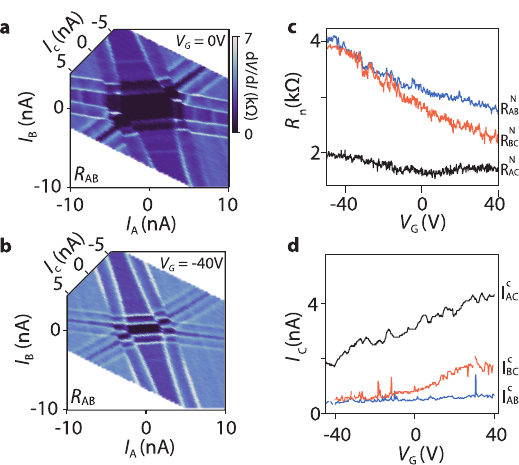}
\vspace{-0.5cm}
\caption{\textbf{Gate- and phase dependence of Multi-terminal Josephson device.} \textbf{a-b}, Differential resistance, $R_\mathrm{AB}$ at back-gate potentials, $V_\mathrm{G}$, of $0\,\mathrm V$ and $-40 \, \mathrm V$, respectively. \textbf{c-d}, the gate dependence of the normal state resistances and critical currents between the three pairs of terminals. }
\vspace{-0.5cm}
\label{fig4}
\end{figure}

An important motivation for using semiconductor nanowire-based JJs is that transport is governed by a few strongly transmitting channels rather than many channels of low transmission and that the number of channels can be controlled by gating\cite{Doh2005,spanton2017current}. Gate tunability and quantized conductance have been reported for single-nanowire junctions based on InAsSb\cite{sestoft:2018,khan2020highly} Gate-control of InAsSb nanowires have previously been reported, however, the variation in bandgap and charge carrier distribution with ternary alloy fraction affects the gatability, which may result in a weak response to gating \cite{sestoft:2018}. In Fig.\ \ref{fig4}, we confirm the gatability of the four-terminal junction, focusing on the effect of the global back-gate. Figure \ref{fig4} (a,b) show the current dependence of $R_\mathrm{AB}$ for two values of the back-gate potential $V_{\mathrm{G}} = 0 \, \mathrm V$ and $V_{\mathrm{bg}} = -40 \, \mathrm V$. Additional data on current map gate dependence is presented in supporting information S8. The InAsSb acts as an $n$-type semiconductor and the width of all branches of critical currents decreases with decreasing $V_{\mathrm{G}}$ while the normal state differential resistance increases. To quantify the gate dependence, measurements along line cuts of each supercurrent branch along paths indicated in Fig.\ \ref{fig3}(d-f) were performed continuously vs. $V_\mathrm{G}$ and the extracted gate-dependence of $I^C_\mathrm{AB}$, $I^C_\mathrm{AC}$ and $I^C_\mathrm{BC}$ and the normal-state values of $R_\mathrm{AB}$, $R_\mathrm{AC}$, and $R_\mathrm{BC}$ extracted outside the superconducting branches, are shown in Fig.\ \ref{fig4}(c,d). For all branches, the critical current(normal state resistance) increases(decreases) with increasing $V_\mathrm G$ thus confirming the 
gatability of the nanowire cross similar to conventional nanowires JJ.
%

\section{Conclusion}

In conclusion, we have presented a new platform of InAsSb-Al nanocross for multiterminal JJ devices. Initially, we presented the growth of these ternary junctions with Al hybridization using UHV MBE. We proposed an understanding of the growth and detailed merging mechanism in the presence of excessive arsenic. We also presented a rigorous structural investigation of nanocrosses showing overall crystal phase and direction are maintained for individual legs before and after merging. Further atomic scale analysis enabled us to reveal how the crystal formation happened in the central region by creating twin boundaries and maintaining crystal system. Finally, we presented the realization of a four-terminal JJ using these ternary materials platform. We demonstrated the supercurrents flowing in the junction by means of DC transport, showing that each terminal contributes to the supercurrent. This is further supported by the flux dependence of the device. Next, we showed the gate behavior of the device, which is critical to facilitate tuning to the single channel regime.


\section{Methods}

\subsection{Hybrid Crystal Growth}

In this project, InAsSb NCs were grown with InAs stem from (111)B trenches. The V-shaped (111)B trenches were created on InAs (100) substrate through wet etching. Afterward, e-beam lithography was used for the control position of Au catalyst particles on the sidewall of the trenches. Physical vapor deposition (PVD) was used to deposit Au film. The detailed substrate fabrication process is reported in \cite{khan2021multiterminal, khan2020highly}. We used an ultra-high vacuum growth cluster for NC growth. The cluster consists of a Vecco Gen II MBE system mainly used for III-V materials growth and an electron gun-assisted physical vapor deposition (PVD) system mainly used for superconducting thin film growth. Initially, InAsSb NC growth was performed in the MBE chamber. The III-V sources were preheated, and each source temperature was stabilized before the growth to confirm uniform incoming flux. The optimized cell temperature for this growth: In- (865/815)$^{\circ}$C, As- (400/345)$^{\circ}$C and Sb- (680/500)$^{\circ}$C. The InAs segment is grown for 12 min maintaining As/In ratio $\sim$ 9.78. Later, the Sb flux is introduced in the system with optimal Sb/In flux ratio $\sim$ 4.30 and In growth rate 0.289 ML/s, which is expected to provide a high aspect ratio and uniform morphology of the Sb based NWs on the trenches. Simultaneously, As to In ratio is lowered to 2.04 by tuning the As valve opening, which helps us to achieve the right composition in the InAsSb NW. Standard InAsSb segment is grown for 30-35 min providing the length of 4-4.5 $\mu$m long NW with diameter varies from 100-150 nm. The diameter variance depends on the catalyst size, which is usually varied from one growth section to another. Post NW radial growth is performed with $\mathrm{As_2}$ (contrary to the NW growth, which is performed with $\mathrm{As_4}$) by breaking As molecules in the cracker at high temperature. For this, after the InAsSb growth, the As source temperature is elevated to 700 $^{\circ}$C and stabilized for 15 min, and then 10 min $\mathrm{As_2}$ assisted overgrowth is performed confirming all the intended NWs are merged properly. Once the NC growth is done, either the sample holder was cooled down to grow Al within the same chamber or the sample was transferred to the PVD chamber to deposit Al on the selected facets. Low-temperature Al growth is performed following the recipe discussed in Ref \cite{khan2020highly}.

\subsection{Device fabrication}
Individual NCs were transferred using a piezo-contolled micro-manipultator from the growth substrate onto a substrate of degenerately doped-Si capped with $200\, \mathrm{nm}$ SiO$_x$. Ohmic contacts and sidegates were defined by electron beam lithography, and $200 \, \mathrm{nm}$ of Al was deposited following a brief \emph{in-situ} Kaufmann milling to remove the native oxide of the Al-coated crosses. In a subsequent lithography step a $400 \, \mathrm{nm}$ local etch window was opened around the intersection of the nanowires and the Al was removed by selective etching in Transene D.



\section{Acknowledgements}
SAK acknowledges funding from Danish Agency for Higher Education and Science, DIANA Quantum Center (DIANAQ), The Danish Chips Competence Center (DKCCC), Innovation Fund Denmark (IFD) projects: DeQD, QLIGHT, European Union Horizon 2020 research and innovation program under the Marie Sk\l{}odowska-Curie Grant No. 722176 (INDEED), European Metrology projects: MetSuperQ (Project No. 23FUN08), NoQTeS (Project No. 23NRM04), DINAMO (Project No. 24DIT02). Materials growth of this project is performed at NBI Molecular Beam Epitaxy System and authors acknowledge C. B. S\o{}rensen for all the maintenance of the system. JJ acknowledges NanoLund for funding. ICN2 acknowledges funding from Generalitat de Catalunya 2021SGR00457. This study is part of the Advanced Materials programme and was supported by MCIN with funding from European Union NextGenerationEU (PRTR-C17.I1) and by Generalitat de Catalunya (In-CAEM Project). We acknowledge support from CSIC Interdisciplinary Thematic Platform (PTI+) on Quantum Technologies (PTI-QTEP+). This research work has been funded by the European Commission – NextGenerationEU (Regulation EU 2020/2094), through CSIC's Quantum Technologies Platform (QTEP). ICN2 is supported by the Severo Ochoa program from Spanish MCIN / AEI (Grant No.: CEX2021-001214-S) and is funded by the CERCA Programme / Generalitat de Catalunya. Authors acknowledge the use of instrumentation as well as the technical advice provided by the Joint Electron Microscopy Center at ALBA (JEMCA). ICN2 acknowledges funding from Grant IU16-014206 (METCAM-FIB) funded by the European Union through the European Regional Development Fund (ERDF), with the support of the Ministry of Research and Universities, Generalitat de Catalunya. ICN2 is founding member of e-DREAM \cite{ciancio2022dream}.


\section{Author Contributions}
 
 S.A.K, L.S and T.S.J conceived the idea, experimental plans. S.A.K and L.S performed substrate fabrication for the growth. S.A.K performed the materials growth, morphological characterization, and related data analysis. S.M.S performed the (S)TEM related experiments, analyzed TEM and EELS data and prepared the atomic models. S.A.K, and J.J proposed the growth mechanisms based on the experimental analysis. L.S, D.O and D.J.C performed the device fabrication, electrical measurements and related data analysis. S.A.K, L.S, T.J and T.S.J prepared the manuscript with the contribution from all other authors. T.S.J, P.K and J.A supervised the project.     

\section{Competing financial interests}
The authors declare no competing financial interests.

\section{Supplementary Information}

S1 - Energy dispersive x-ray spectroscopy (EDX) to analyze the NC composition; S2 - Additional Analysis in the Edge of the junction area; S3 - Additional Analysis in the Overgrown region of the junction area; S4 - Compression in the overgrown area of the Nanocross Junction; S5 - Additional data on other devices; S6 - Differential resistance maps with different sweep directions; S7 - Additional data for flux-dependence.

\bibliography{ref,RefsTSJ}

\begin{thebibliography}{37}%
\makeatletter
\providecommand \@ifxundefined [1]{%
 \@ifx{#1\undefined}
}%
\providecommand \@ifnum [1]{%
 \ifnum #1\expandafter \@firstoftwo
 \else \expandafter \@secondoftwo
 \fi
}%
\providecommand \@ifx [1]{%
 \ifx #1\expandafter \@firstoftwo
 \else \expandafter \@secondoftwo
 \fi
}%
\providecommand \natexlab [1]{#1}%
\providecommand \enquote  [1]{``#1''}%
\providecommand \bibnamefont  [1]{#1}%
\providecommand \bibfnamefont [1]{#1}%
\providecommand \citenamefont [1]{#1}%
\providecommand \href@noop [0]{\@secondoftwo}%
\providecommand \href [0]{\begingroup \@sanitize@url \@href}%
\providecommand \@href[1]{\@@startlink{#1}\@@href}%
\providecommand \@@href[1]{\endgroup#1\@@endlink}%
\providecommand \@sanitize@url [0]{\catcode `\\12\catcode `\$12\catcode
  `\&12\catcode `\#12\catcode `\^12\catcode `\_12\catcode `\%12\relax}%
\providecommand \@@startlink[1]{}%
\providecommand \@@endlink[0]{}%
\providecommand \url  [0]{\begingroup\@sanitize@url \@url }%
\providecommand \@url [1]{\endgroup\@href {#1}{\urlprefix }}%
\providecommand \urlprefix  [0]{URL }%
\providecommand \Eprint [0]{\href }%
\providecommand \doibase [0]{http://dx.doi.org/}%
\providecommand \selectlanguage [0]{\@gobble}%
\providecommand \bibinfo  [0]{\@secondoftwo}%
\providecommand \bibfield  [0]{\@secondoftwo}%
\providecommand \translation [1]{[#1]}%
\providecommand \BibitemOpen [0]{}%
\providecommand \bibitemStop [0]{}%
\providecommand \bibitemNoStop [0]{.\EOS\space}%
\providecommand \EOS [0]{\spacefactor3000\relax}%
\providecommand \BibitemShut  [1]{\csname bibitem#1\endcsname}%
\let\auto@bib@innerbib\@empty
\bibitem [{\citenamefont {Riwar}\ \emph {et~al.}(2016)\citenamefont {Riwar},
  \citenamefont {Houzet}, \citenamefont {Meyer},\ and\ \citenamefont
  {Nazarov}}]{riwar2016multi}%
  \BibitemOpen
  \bibfield  {author} {\bibinfo {author} {\bibfnamefont {R.-P.}\ \bibnamefont
  {Riwar}}, \bibinfo {author} {\bibfnamefont {M.}~\bibnamefont {Houzet}},
  \bibinfo {author} {\bibfnamefont {J.~S.}\ \bibnamefont {Meyer}}, \ and\
  \bibinfo {author} {\bibfnamefont {Y.~V.}\ \bibnamefont {Nazarov}},\
  }\href@noop {} {\bibfield  {journal} {\bibinfo  {journal} {Nature
  communications}\ }\textbf {\bibinfo {volume} {7}},\ \bibinfo {pages} {11167}
  (\bibinfo {year} {2016})}\BibitemShut {NoStop}%
\bibitem [{\citenamefont {Pankratova}\ \emph {et~al.}(2020)\citenamefont
  {Pankratova}, \citenamefont {Lee}, \citenamefont {Kuzmin}, \citenamefont
  {Wickramasinghe}, \citenamefont {Mayer}, \citenamefont {Yuan}, \citenamefont
  {Vavilov}, \citenamefont {Shabani},\ and\ \citenamefont
  {Manucharyan}}]{pankratova2020multiterminal}%
  \BibitemOpen
  \bibfield  {author} {\bibinfo {author} {\bibfnamefont {N.}~\bibnamefont
  {Pankratova}}, \bibinfo {author} {\bibfnamefont {H.}~\bibnamefont {Lee}},
  \bibinfo {author} {\bibfnamefont {R.}~\bibnamefont {Kuzmin}}, \bibinfo
  {author} {\bibfnamefont {K.}~\bibnamefont {Wickramasinghe}}, \bibinfo
  {author} {\bibfnamefont {W.}~\bibnamefont {Mayer}}, \bibinfo {author}
  {\bibfnamefont {J.}~\bibnamefont {Yuan}}, \bibinfo {author} {\bibfnamefont
  {M.~G.}\ \bibnamefont {Vavilov}}, \bibinfo {author} {\bibfnamefont
  {J.}~\bibnamefont {Shabani}}, \ and\ \bibinfo {author} {\bibfnamefont
  {V.~E.}\ \bibnamefont {Manucharyan}},\ }\href@noop {} {\bibfield  {journal}
  {\bibinfo  {journal} {Physical Review X}\ }\textbf {\bibinfo {volume} {10}},\
  \bibinfo {pages} {031051} (\bibinfo {year} {2020})}\BibitemShut {NoStop}%
\bibitem [{\citenamefont {Draelos}\ \emph
  {et~al.}(2019{\natexlab{a}})\citenamefont {Draelos}, \citenamefont {Wei},
  \citenamefont {Seredinski}, \citenamefont {Li}, \citenamefont {Mehta},
  \citenamefont {Watanabe}, \citenamefont {Taniguchi}, \citenamefont
  {Borzenets}, \citenamefont {Amet},\ and\ \citenamefont
  {Finkelstein}}]{draelos2019supercurrent}%
  \BibitemOpen
  \bibfield  {author} {\bibinfo {author} {\bibfnamefont {A.~W.}\ \bibnamefont
  {Draelos}}, \bibinfo {author} {\bibfnamefont {M.-T.}\ \bibnamefont {Wei}},
  \bibinfo {author} {\bibfnamefont {A.}~\bibnamefont {Seredinski}}, \bibinfo
  {author} {\bibfnamefont {H.}~\bibnamefont {Li}}, \bibinfo {author}
  {\bibfnamefont {Y.}~\bibnamefont {Mehta}}, \bibinfo {author} {\bibfnamefont
  {K.}~\bibnamefont {Watanabe}}, \bibinfo {author} {\bibfnamefont
  {T.}~\bibnamefont {Taniguchi}}, \bibinfo {author} {\bibfnamefont {I.~V.}\
  \bibnamefont {Borzenets}}, \bibinfo {author} {\bibfnamefont {F.}~\bibnamefont
  {Amet}}, \ and\ \bibinfo {author} {\bibfnamefont {G.}~\bibnamefont
  {Finkelstein}},\ }\href@noop {} {\bibfield  {journal} {\bibinfo  {journal}
  {Nano letters}\ }\textbf {\bibinfo {volume} {19}},\ \bibinfo {pages} {1039}
  (\bibinfo {year} {2019}{\natexlab{a}})}\BibitemShut {NoStop}%
\bibitem [{\citenamefont {Zhang}\ \emph {et~al.}(2023)\citenamefont {Zhang},
  \citenamefont {Rashid}, \citenamefont {Ahari}, \citenamefont {Zhang},
  \citenamefont {Ananthanarayanan}, \citenamefont {Xiao}, \citenamefont
  {De~Coster}, \citenamefont {Gilbert}, \citenamefont {Samarth},\ and\
  \citenamefont {Kayyalha}}]{zhang2023andreev}%
  \BibitemOpen
  \bibfield  {author} {\bibinfo {author} {\bibfnamefont {F.}~\bibnamefont
  {Zhang}}, \bibinfo {author} {\bibfnamefont {A.~S.}\ \bibnamefont {Rashid}},
  \bibinfo {author} {\bibfnamefont {M.~T.}\ \bibnamefont {Ahari}}, \bibinfo
  {author} {\bibfnamefont {W.}~\bibnamefont {Zhang}}, \bibinfo {author}
  {\bibfnamefont {K.~M.}\ \bibnamefont {Ananthanarayanan}}, \bibinfo {author}
  {\bibfnamefont {R.}~\bibnamefont {Xiao}}, \bibinfo {author} {\bibfnamefont
  {G.~J.}\ \bibnamefont {De~Coster}}, \bibinfo {author} {\bibfnamefont {M.~J.}\
  \bibnamefont {Gilbert}}, \bibinfo {author} {\bibfnamefont {N.}~\bibnamefont
  {Samarth}}, \ and\ \bibinfo {author} {\bibfnamefont {M.}~\bibnamefont
  {Kayyalha}},\ }\href@noop {} {\bibfield  {journal} {\bibinfo  {journal}
  {Physical Review B}\ }\textbf {\bibinfo {volume} {107}},\ \bibinfo {pages}
  {L140503} (\bibinfo {year} {2023})}\BibitemShut {NoStop}%
\bibitem [{\citenamefont {Schiela}\ \emph {et~al.}(2024)\citenamefont
  {Schiela}, \citenamefont {Yu},\ and\ \citenamefont
  {Shabani}}]{schiela2024progress}%
  \BibitemOpen
  \bibfield  {author} {\bibinfo {author} {\bibfnamefont {W.~F.}\ \bibnamefont
  {Schiela}}, \bibinfo {author} {\bibfnamefont {P.}~\bibnamefont {Yu}}, \ and\
  \bibinfo {author} {\bibfnamefont {J.}~\bibnamefont {Shabani}},\ }\href@noop
  {} {\bibfield  {journal} {\bibinfo  {journal} {PRX Quantum}\ }\textbf
  {\bibinfo {volume} {5}},\ \bibinfo {pages} {030102} (\bibinfo {year}
  {2024})}\BibitemShut {NoStop}%
\bibitem [{\citenamefont {Zhang}\ \emph {et~al.}(2024)\citenamefont {Zhang},
  \citenamefont {Zarassi}, \citenamefont {Jarjat}, \citenamefont {Van~de
  Sande}, \citenamefont {Pendharkar}, \citenamefont {Lee}, \citenamefont
  {Dempsey}, \citenamefont {McFadden}, \citenamefont {Harrington},
  \citenamefont {Dong} \emph {et~al.}}]{zhang2024large}%
  \BibitemOpen
  \bibfield  {author} {\bibinfo {author} {\bibfnamefont {P.}~\bibnamefont
  {Zhang}}, \bibinfo {author} {\bibfnamefont {A.}~\bibnamefont {Zarassi}},
  \bibinfo {author} {\bibfnamefont {L.}~\bibnamefont {Jarjat}}, \bibinfo
  {author} {\bibfnamefont {V.}~\bibnamefont {Van~de Sande}}, \bibinfo {author}
  {\bibfnamefont {M.}~\bibnamefont {Pendharkar}}, \bibinfo {author}
  {\bibfnamefont {J.}~\bibnamefont {Lee}}, \bibinfo {author} {\bibfnamefont
  {C.~P.}\ \bibnamefont {Dempsey}}, \bibinfo {author} {\bibfnamefont
  {A.}~\bibnamefont {McFadden}}, \bibinfo {author} {\bibfnamefont {S.~D.}\
  \bibnamefont {Harrington}}, \bibinfo {author} {\bibfnamefont {J.~T.}\
  \bibnamefont {Dong}},  \emph {et~al.},\ }\href@noop {} {\bibfield  {journal}
  {\bibinfo  {journal} {SciPost Physics}\ }\textbf {\bibinfo {volume} {16}},\
  \bibinfo {pages} {030} (\bibinfo {year} {2024})}\BibitemShut {NoStop}%
\bibitem [{\citenamefont {Khan}\ \emph {et~al.}(2020)\citenamefont {Khan},
  \citenamefont {Lampadaris}, \citenamefont {Cui}, \citenamefont {Stampfer},
  \citenamefont {Liu}, \citenamefont {Pauka}, \citenamefont {Cachaza},
  \citenamefont {Fiordaliso}, \citenamefont {Kang}, \citenamefont {Korneychuk}
  \emph {et~al.}}]{khan2020highly}%
  \BibitemOpen
  \bibfield  {author} {\bibinfo {author} {\bibfnamefont {S.~A.}\ \bibnamefont
  {Khan}}, \bibinfo {author} {\bibfnamefont {C.}~\bibnamefont {Lampadaris}},
  \bibinfo {author} {\bibfnamefont {A.}~\bibnamefont {Cui}}, \bibinfo {author}
  {\bibfnamefont {L.}~\bibnamefont {Stampfer}}, \bibinfo {author}
  {\bibfnamefont {Y.}~\bibnamefont {Liu}}, \bibinfo {author} {\bibfnamefont
  {S.~J.}\ \bibnamefont {Pauka}}, \bibinfo {author} {\bibfnamefont {M.~E.}\
  \bibnamefont {Cachaza}}, \bibinfo {author} {\bibfnamefont {E.~M.}\
  \bibnamefont {Fiordaliso}}, \bibinfo {author} {\bibfnamefont {J.-H.}\
  \bibnamefont {Kang}}, \bibinfo {author} {\bibfnamefont {S.}~\bibnamefont
  {Korneychuk}},  \emph {et~al.},\ }\href@noop {} {\bibfield  {journal}
  {\bibinfo  {journal} {ACS nano}\ }\textbf {\bibinfo {volume} {14}},\ \bibinfo
  {pages} {14605} (\bibinfo {year} {2020})}\BibitemShut {NoStop}%
\bibitem [{\citenamefont {Carrad}\ \emph {et~al.}(2020)\citenamefont {Carrad},
  \citenamefont {Bjergfelt}, \citenamefont {Kanne}, \citenamefont {Aagesen},
  \citenamefont {Krizek}, \citenamefont {Fiordaliso}, \citenamefont {Johnson},
  \citenamefont {Nyg{\aa}rd},\ and\ \citenamefont
  {Jespersen}}]{carrad2020shadow}%
  \BibitemOpen
  \bibfield  {author} {\bibinfo {author} {\bibfnamefont {D.~J.}\ \bibnamefont
  {Carrad}}, \bibinfo {author} {\bibfnamefont {M.}~\bibnamefont {Bjergfelt}},
  \bibinfo {author} {\bibfnamefont {T.}~\bibnamefont {Kanne}}, \bibinfo
  {author} {\bibfnamefont {M.}~\bibnamefont {Aagesen}}, \bibinfo {author}
  {\bibfnamefont {F.}~\bibnamefont {Krizek}}, \bibinfo {author} {\bibfnamefont
  {E.~M.}\ \bibnamefont {Fiordaliso}}, \bibinfo {author} {\bibfnamefont
  {E.}~\bibnamefont {Johnson}}, \bibinfo {author} {\bibfnamefont
  {J.}~\bibnamefont {Nyg{\aa}rd}}, \ and\ \bibinfo {author} {\bibfnamefont
  {T.~S.}\ \bibnamefont {Jespersen}},\ }\href@noop {} {\bibfield  {journal}
  {\bibinfo  {journal} {Advanced Materials}\ }\textbf {\bibinfo {volume}
  {32}},\ \bibinfo {pages} {1908411} (\bibinfo {year} {2020})}\BibitemShut
  {NoStop}%
\bibitem [{\citenamefont {Paudel}\ \emph {et~al.}(2025)\citenamefont {Paudel},
  \citenamefont {Smith},\ and\ \citenamefont {Stanescu}}]{paudel2025disorder}%
  \BibitemOpen
  \bibfield  {author} {\bibinfo {author} {\bibfnamefont {P.~P.}\ \bibnamefont
  {Paudel}}, \bibinfo {author} {\bibfnamefont {N.~O.}\ \bibnamefont {Smith}}, \
  and\ \bibinfo {author} {\bibfnamefont {T.~D.}\ \bibnamefont {Stanescu}},\
  }\href@noop {} {\bibfield  {journal} {\bibinfo  {journal} {Physical Review
  Applied}\ }\textbf {\bibinfo {volume} {23}},\ \bibinfo {pages} {034068}
  (\bibinfo {year} {2025})}\BibitemShut {NoStop}%
\bibitem [{\citenamefont {Khan}\ \emph {et~al.}(2021)\citenamefont {Khan},
  \citenamefont {Stampfer}, \citenamefont {Mutas}, \citenamefont {Kang},
  \citenamefont {Krogstrup},\ and\ \citenamefont
  {Jespersen}}]{khan2021multiterminal}%
  \BibitemOpen
  \bibfield  {author} {\bibinfo {author} {\bibfnamefont {S.~A.}\ \bibnamefont
  {Khan}}, \bibinfo {author} {\bibfnamefont {L.}~\bibnamefont {Stampfer}},
  \bibinfo {author} {\bibfnamefont {T.}~\bibnamefont {Mutas}}, \bibinfo
  {author} {\bibfnamefont {J.-H.}\ \bibnamefont {Kang}}, \bibinfo {author}
  {\bibfnamefont {P.}~\bibnamefont {Krogstrup}}, \ and\ \bibinfo {author}
  {\bibfnamefont {T.~S.}\ \bibnamefont {Jespersen}},\ }\href@noop {} {\bibfield
   {journal} {\bibinfo  {journal} {Advanced Materials}\ ,\ \bibinfo {pages}
  {2100078}} (\bibinfo {year} {2021})}\BibitemShut {NoStop}%
\bibitem [{\citenamefont {Rossi}\ \emph {et~al.}(2024)\citenamefont {Rossi},
  \citenamefont {van Schijndel}, \citenamefont {Lueb}, \citenamefont {Badawy},
  \citenamefont {Jung}, \citenamefont {Peeters}, \citenamefont {K{\"o}lling},
  \citenamefont {Moutanabbir}, \citenamefont {Verheijen},\ and\ \citenamefont
  {Bakkers}}]{rossi2024stemless}%
  \BibitemOpen
  \bibfield  {author} {\bibinfo {author} {\bibfnamefont {M.}~\bibnamefont
  {Rossi}}, \bibinfo {author} {\bibfnamefont {T.~A.}\ \bibnamefont {van
  Schijndel}}, \bibinfo {author} {\bibfnamefont {P.}~\bibnamefont {Lueb}},
  \bibinfo {author} {\bibfnamefont {G.}~\bibnamefont {Badawy}}, \bibinfo
  {author} {\bibfnamefont {J.}~\bibnamefont {Jung}}, \bibinfo {author}
  {\bibfnamefont {W.~H.}\ \bibnamefont {Peeters}}, \bibinfo {author}
  {\bibfnamefont {S.}~\bibnamefont {K{\"o}lling}}, \bibinfo {author}
  {\bibfnamefont {O.}~\bibnamefont {Moutanabbir}}, \bibinfo {author}
  {\bibfnamefont {M.~A.}\ \bibnamefont {Verheijen}}, \ and\ \bibinfo {author}
  {\bibfnamefont {E.~P.}\ \bibnamefont {Bakkers}},\ }\href@noop {} {\bibfield
  {journal} {\bibinfo  {journal} {Nanotechnology}\ }\textbf {\bibinfo {volume}
  {35}},\ \bibinfo {pages} {415602} (\bibinfo {year} {2024})}\BibitemShut
  {NoStop}%
\bibitem [{\citenamefont {Krizek}\ \emph {et~al.}(2018)\citenamefont {Krizek},
  \citenamefont {Sestoft}, \citenamefont {Aseev}, \citenamefont
  {Marti-Sanchez}, \citenamefont {Vaitiek\ifmmode~\dot{e}\else \.{e}\fi{}nas},
  \citenamefont {Casparis}, \citenamefont {Khan}, \citenamefont {Liu},
  \citenamefont {Stankevi\ifmmode~\check{c}\else \v{c}\fi{}}, \citenamefont
  {Whiticar}, \citenamefont {Fursina}, \citenamefont {Boekhout}, \citenamefont
  {Koops}, \citenamefont {Uccelli}, \citenamefont {Kouwenhoven}, \citenamefont
  {Marcus}, \citenamefont {Arbiol},\ and\ \citenamefont
  {Krogstrup}}]{KrizekSAG2018}%
  \BibitemOpen
  \bibfield  {author} {\bibinfo {author} {\bibfnamefont {F.}~\bibnamefont
  {Krizek}}, \bibinfo {author} {\bibfnamefont {J.~E.}\ \bibnamefont {Sestoft}},
  \bibinfo {author} {\bibfnamefont {P.}~\bibnamefont {Aseev}}, \bibinfo
  {author} {\bibfnamefont {S.}~\bibnamefont {Marti-Sanchez}}, \bibinfo {author}
  {\bibfnamefont {S.}~\bibnamefont {Vaitiek\ifmmode~\dot{e}\else
  \.{e}\fi{}nas}}, \bibinfo {author} {\bibfnamefont {L.}~\bibnamefont
  {Casparis}}, \bibinfo {author} {\bibfnamefont {S.~A.}\ \bibnamefont {Khan}},
  \bibinfo {author} {\bibfnamefont {Y.}~\bibnamefont {Liu}}, \bibinfo {author}
  {\bibfnamefont {T.~c.~v.}\ \bibnamefont {Stankevi\ifmmode~\check{c}\else
  \v{c}\fi{}}}, \bibinfo {author} {\bibfnamefont {A.~M.}\ \bibnamefont
  {Whiticar}}, \bibinfo {author} {\bibfnamefont {A.}~\bibnamefont {Fursina}},
  \bibinfo {author} {\bibfnamefont {F.}~\bibnamefont {Boekhout}}, \bibinfo
  {author} {\bibfnamefont {R.}~\bibnamefont {Koops}}, \bibinfo {author}
  {\bibfnamefont {E.}~\bibnamefont {Uccelli}}, \bibinfo {author} {\bibfnamefont
  {L.~P.}\ \bibnamefont {Kouwenhoven}}, \bibinfo {author} {\bibfnamefont
  {C.~M.}\ \bibnamefont {Marcus}}, \bibinfo {author} {\bibfnamefont
  {J.}~\bibnamefont {Arbiol}}, \ and\ \bibinfo {author} {\bibfnamefont
  {P.}~\bibnamefont {Krogstrup}},\ }\href@noop {} {\bibfield  {journal}
  {\bibinfo  {journal} {Physical Review. Materials}\ }\textbf {\bibinfo
  {volume} {2}},\ \bibinfo {pages} {093401} (\bibinfo {year}
  {2018})}\BibitemShut {NoStop}%
\bibitem [{\citenamefont {Friedl}\ \emph {et~al.}(2018)\citenamefont {Friedl},
  \citenamefont {Cerveny}, \citenamefont {Weigele}, \citenamefont
  {T\"{u}t\"{u}nc\"{u}oglu}, \citenamefont {Mart{\'i}-S{\'a}nchez},
  \citenamefont {Huang}, \citenamefont {Patlatiuk}, \citenamefont {Potts},
  \citenamefont {Sun}, \citenamefont {Hill} \emph
  {et~al.}}]{friedl2018template}%
  \BibitemOpen
  \bibfield  {author} {\bibinfo {author} {\bibfnamefont {M.}~\bibnamefont
  {Friedl}}, \bibinfo {author} {\bibfnamefont {K.}~\bibnamefont {Cerveny}},
  \bibinfo {author} {\bibfnamefont {P.}~\bibnamefont {Weigele}}, \bibinfo
  {author} {\bibfnamefont {G.}~\bibnamefont {T\"{u}t\"{u}nc\"{u}oglu}},
  \bibinfo {author} {\bibfnamefont {S.}~\bibnamefont {Mart{\'i}-S{\'a}nchez}},
  \bibinfo {author} {\bibfnamefont {C.}~\bibnamefont {Huang}}, \bibinfo
  {author} {\bibfnamefont {T.}~\bibnamefont {Patlatiuk}}, \bibinfo {author}
  {\bibfnamefont {H.}~\bibnamefont {Potts}}, \bibinfo {author} {\bibfnamefont
  {Z.}~\bibnamefont {Sun}}, \bibinfo {author} {\bibfnamefont {M.~O.}\
  \bibnamefont {Hill}},  \emph {et~al.},\ }\href@noop {} {\bibfield  {journal}
  {\bibinfo  {journal} {Nano Lett.}\ }\textbf {\bibinfo {volume} {18}},\
  \bibinfo {pages} {2666} (\bibinfo {year} {2018})}\BibitemShut {NoStop}%
\bibitem [{\citenamefont {Aseev}\ \emph {et~al.}(2019)\citenamefont {Aseev},
  \citenamefont {Wang}, \citenamefont {Binci}, \citenamefont {Singh},
  \citenamefont {Mart{\'i}-S{\'a}nchez}, \citenamefont {Botifoll},
  \citenamefont {Stek}, \citenamefont {Bordin}, \citenamefont {Watson},
  \citenamefont {Boekhout} \emph {et~al.}}]{aseev2019ballistic}%
  \BibitemOpen
  \bibfield  {author} {\bibinfo {author} {\bibfnamefont {P.}~\bibnamefont
  {Aseev}}, \bibinfo {author} {\bibfnamefont {G.}~\bibnamefont {Wang}},
  \bibinfo {author} {\bibfnamefont {L.}~\bibnamefont {Binci}}, \bibinfo
  {author} {\bibfnamefont {A.}~\bibnamefont {Singh}}, \bibinfo {author}
  {\bibfnamefont {S.}~\bibnamefont {Mart{\'i}-S{\'a}nchez}}, \bibinfo {author}
  {\bibfnamefont {M.}~\bibnamefont {Botifoll}}, \bibinfo {author}
  {\bibfnamefont {L.~J.}\ \bibnamefont {Stek}}, \bibinfo {author}
  {\bibfnamefont {A.}~\bibnamefont {Bordin}}, \bibinfo {author} {\bibfnamefont
  {J.~D.}\ \bibnamefont {Watson}}, \bibinfo {author} {\bibfnamefont
  {F.}~\bibnamefont {Boekhout}},  \emph {et~al.},\ }\href@noop {} {\bibfield
  {journal} {\bibinfo  {journal} {Nano letters}\ }\textbf {\bibinfo {volume}
  {19}},\ \bibinfo {pages} {9102} (\bibinfo {year} {2019})}\BibitemShut
  {NoStop}%
\bibitem [{\citenamefont {Liu}\ \emph {et~al.}(2019)\citenamefont {Liu},
  \citenamefont {Vaitiekenas}, \citenamefont {Mart{\'i}-S{\'a}nchez},
  \citenamefont {Koch}, \citenamefont {Hart}, \citenamefont {Cui},
  \citenamefont {Kanne}, \citenamefont {Khan}, \citenamefont {Tanta},
  \citenamefont {Upadhyay} \emph {et~al.}}]{liu2019semiconductor}%
  \BibitemOpen
  \bibfield  {author} {\bibinfo {author} {\bibfnamefont {Y.}~\bibnamefont
  {Liu}}, \bibinfo {author} {\bibfnamefont {S.}~\bibnamefont {Vaitiekenas}},
  \bibinfo {author} {\bibfnamefont {S.}~\bibnamefont {Mart{\'i}-S{\'a}nchez}},
  \bibinfo {author} {\bibfnamefont {C.}~\bibnamefont {Koch}}, \bibinfo {author}
  {\bibfnamefont {S.}~\bibnamefont {Hart}}, \bibinfo {author} {\bibfnamefont
  {Z.}~\bibnamefont {Cui}}, \bibinfo {author} {\bibfnamefont {T.}~\bibnamefont
  {Kanne}}, \bibinfo {author} {\bibfnamefont {S.~A.}\ \bibnamefont {Khan}},
  \bibinfo {author} {\bibfnamefont {R.}~\bibnamefont {Tanta}}, \bibinfo
  {author} {\bibfnamefont {S.}~\bibnamefont {Upadhyay}},  \emph {et~al.},\
  }\href@noop {} {\bibfield  {journal} {\bibinfo  {journal} {Nano Letters}\
  }\textbf {\bibinfo {volume} {20}},\ \bibinfo {pages} {456} (\bibinfo {year}
  {2019})}\BibitemShut {NoStop}%
\bibitem [{\citenamefont {Zhang}\ \emph {et~al.}(2022)\citenamefont {Zhang},
  \citenamefont {Wu}, \citenamefont {Chen}, \citenamefont {Khan}, \citenamefont
  {Krogstrup}, \citenamefont {Pekker},\ and\ \citenamefont
  {Frolov}}]{zhang2022signatures}%
  \BibitemOpen
  \bibfield  {author} {\bibinfo {author} {\bibfnamefont {P.}~\bibnamefont
  {Zhang}}, \bibinfo {author} {\bibfnamefont {H.}~\bibnamefont {Wu}}, \bibinfo
  {author} {\bibfnamefont {J.}~\bibnamefont {Chen}}, \bibinfo {author}
  {\bibfnamefont {S.~A.}\ \bibnamefont {Khan}}, \bibinfo {author}
  {\bibfnamefont {P.}~\bibnamefont {Krogstrup}}, \bibinfo {author}
  {\bibfnamefont {D.}~\bibnamefont {Pekker}}, \ and\ \bibinfo {author}
  {\bibfnamefont {S.~M.}\ \bibnamefont {Frolov}},\ }\href@noop {} {\bibfield
  {journal} {\bibinfo  {journal} {Physical Review Letters}\ }\textbf {\bibinfo
  {volume} {128}},\ \bibinfo {pages} {046801} (\bibinfo {year}
  {2022})}\BibitemShut {NoStop}%
\bibitem [{\citenamefont {Khan}\ \emph {et~al.}(2023)\citenamefont {Khan},
  \citenamefont {Mart{\'\i}-S{\'a}nchez}, \citenamefont {Olsteins},
  \citenamefont {Lampadaris}, \citenamefont {Carrad}, \citenamefont {Liu},
  \citenamefont {Qui{\~n}ones}, \citenamefont {Chiara~Spadaro}, \citenamefont
  {Sand~Jespersen}, \citenamefont {Krogstrup} \emph
  {et~al.}}]{khan2023epitaxially}%
  \BibitemOpen
  \bibfield  {author} {\bibinfo {author} {\bibfnamefont {S.~A.}\ \bibnamefont
  {Khan}}, \bibinfo {author} {\bibfnamefont {S.}~\bibnamefont
  {Mart{\'\i}-S{\'a}nchez}}, \bibinfo {author} {\bibfnamefont {D.}~\bibnamefont
  {Olsteins}}, \bibinfo {author} {\bibfnamefont {C.}~\bibnamefont
  {Lampadaris}}, \bibinfo {author} {\bibfnamefont {D.~J.}\ \bibnamefont
  {Carrad}}, \bibinfo {author} {\bibfnamefont {Y.}~\bibnamefont {Liu}},
  \bibinfo {author} {\bibfnamefont {J.}~\bibnamefont {Qui{\~n}ones}}, \bibinfo
  {author} {\bibfnamefont {M.}~\bibnamefont {Chiara~Spadaro}}, \bibinfo
  {author} {\bibfnamefont {T.}~\bibnamefont {Sand~Jespersen}}, \bibinfo
  {author} {\bibfnamefont {P.}~\bibnamefont {Krogstrup}},  \emph {et~al.},\
  }\href@noop {} {\bibfield  {journal} {\bibinfo  {journal} {ACS nano}\
  }\textbf {\bibinfo {volume} {17}},\ \bibinfo {pages} {11794} (\bibinfo {year}
  {2023})}\BibitemShut {NoStop}%
\bibitem [{\citenamefont {Carrad}\ \emph {et~al.}(2022)\citenamefont {Carrad},
  \citenamefont {Stampfer}, \citenamefont {Ol/u{s}teins}, \citenamefont
  {Petersen}, \citenamefont {Khan}, \citenamefont {Krogstrup},\ and\
  \citenamefont {Jespersen}}]{carrad2022photon}%
  \BibitemOpen
  \bibfield  {author} {\bibinfo {author} {\bibfnamefont {D.~J.}\ \bibnamefont
  {Carrad}}, \bibinfo {author} {\bibfnamefont {L.}~\bibnamefont {Stampfer}},
  \bibinfo {author} {\bibfnamefont {D.}~\bibnamefont {Ol/u{s}teins}}, \bibinfo
  {author} {\bibfnamefont {C.~E.~N.}\ \bibnamefont {Petersen}}, \bibinfo
  {author} {\bibfnamefont {S.~A.}\ \bibnamefont {Khan}}, \bibinfo {author}
  {\bibfnamefont {P.}~\bibnamefont {Krogstrup}}, \ and\ \bibinfo {author}
  {\bibfnamefont {T.~S.}\ \bibnamefont {Jespersen}},\ }\href@noop {} {\bibfield
   {journal} {\bibinfo  {journal} {Nano Letters}\ }\textbf {\bibinfo {volume}
  {22}},\ \bibinfo {pages} {6262} (\bibinfo {year} {2022})}\BibitemShut
  {NoStop}%
\bibitem [{\citenamefont {Chen}\ \emph {et~al.}(2023)\citenamefont {Chen},
  \citenamefont {van Driel}, \citenamefont {Lampadaris}, \citenamefont {Khan},
  \citenamefont {Alattallah}, \citenamefont {Zeng}, \citenamefont {Olsson},
  \citenamefont {Dvir}, \citenamefont {Krogstrup},\ and\ \citenamefont
  {Liu}}]{chen2023gate}%
  \BibitemOpen
  \bibfield  {author} {\bibinfo {author} {\bibfnamefont {Y.}~\bibnamefont
  {Chen}}, \bibinfo {author} {\bibfnamefont {D.}~\bibnamefont {van Driel}},
  \bibinfo {author} {\bibfnamefont {C.}~\bibnamefont {Lampadaris}}, \bibinfo
  {author} {\bibfnamefont {S.~A.}\ \bibnamefont {Khan}}, \bibinfo {author}
  {\bibfnamefont {K.}~\bibnamefont {Alattallah}}, \bibinfo {author}
  {\bibfnamefont {L.}~\bibnamefont {Zeng}}, \bibinfo {author} {\bibfnamefont
  {E.}~\bibnamefont {Olsson}}, \bibinfo {author} {\bibfnamefont
  {T.}~\bibnamefont {Dvir}}, \bibinfo {author} {\bibfnamefont {P.}~\bibnamefont
  {Krogstrup}}, \ and\ \bibinfo {author} {\bibfnamefont {Y.}~\bibnamefont
  {Liu}},\ }\href@noop {} {\bibfield  {journal} {\bibinfo  {journal} {Applied
  Physics Letters}\ }\textbf {\bibinfo {volume} {123}} (\bibinfo {year}
  {2023})}\BibitemShut {NoStop}%
\bibitem [{\citenamefont {Stavenga}\ \emph {et~al.}(2023)\citenamefont
  {Stavenga}, \citenamefont {Khan}, \citenamefont {Liu}, \citenamefont
  {Krogstrup},\ and\ \citenamefont {DiCarlo}}]{stavenga2023lower}%
  \BibitemOpen
  \bibfield  {author} {\bibinfo {author} {\bibfnamefont {T.}~\bibnamefont
  {Stavenga}}, \bibinfo {author} {\bibfnamefont {S.}~\bibnamefont {Khan}},
  \bibinfo {author} {\bibfnamefont {Y.}~\bibnamefont {Liu}}, \bibinfo {author}
  {\bibfnamefont {P.}~\bibnamefont {Krogstrup}}, \ and\ \bibinfo {author}
  {\bibfnamefont {L.}~\bibnamefont {DiCarlo}},\ }\href@noop {} {\bibfield
  {journal} {\bibinfo  {journal} {Applied Physics Letters}\ }\textbf {\bibinfo
  {volume} {123}} (\bibinfo {year} {2023})}\BibitemShut {NoStop}%
\bibitem [{\citenamefont {Goswami}\ \emph {et~al.}(2023)\citenamefont
  {Goswami}, \citenamefont {Mudi}, \citenamefont {Dempsey}, \citenamefont
  {Zhang}, \citenamefont {Wu}, \citenamefont {Zhang}, \citenamefont {Mitchell},
  \citenamefont {Lee}, \citenamefont {Frolov},\ and\ \citenamefont
  {Palmstr{\o}m}}]{goswami2023sn}%
  \BibitemOpen
  \bibfield  {author} {\bibinfo {author} {\bibfnamefont {A.}~\bibnamefont
  {Goswami}}, \bibinfo {author} {\bibfnamefont {S.~R.}\ \bibnamefont {Mudi}},
  \bibinfo {author} {\bibfnamefont {C.}~\bibnamefont {Dempsey}}, \bibinfo
  {author} {\bibfnamefont {P.}~\bibnamefont {Zhang}}, \bibinfo {author}
  {\bibfnamefont {H.}~\bibnamefont {Wu}}, \bibinfo {author} {\bibfnamefont
  {B.}~\bibnamefont {Zhang}}, \bibinfo {author} {\bibfnamefont {W.~J.}\
  \bibnamefont {Mitchell}}, \bibinfo {author} {\bibfnamefont {J.~S.}\
  \bibnamefont {Lee}}, \bibinfo {author} {\bibfnamefont {S.~M.}\ \bibnamefont
  {Frolov}}, \ and\ \bibinfo {author} {\bibfnamefont {C.~J.}\ \bibnamefont
  {Palmstr{\o}m}},\ }\href@noop {} {\bibfield  {journal} {\bibinfo  {journal}
  {Nano Letters}\ }\textbf {\bibinfo {volume} {23}},\ \bibinfo {pages} {7311}
  (\bibinfo {year} {2023})}\BibitemShut {NoStop}%
\bibitem [{\citenamefont {Badawy}\ \emph {et~al.}(2023)\citenamefont {Badawy},
  \citenamefont {Verheijen},\ and\ \citenamefont
  {Bakkers}}]{badawy2023tunable}%
  \BibitemOpen
  \bibfield  {author} {\bibinfo {author} {\bibfnamefont {G.}~\bibnamefont
  {Badawy}}, \bibinfo {author} {\bibfnamefont {M.~A.}\ \bibnamefont
  {Verheijen}}, \ and\ \bibinfo {author} {\bibfnamefont {E.~P.}\ \bibnamefont
  {Bakkers}},\ }\href@noop {} {\bibfield  {journal} {\bibinfo  {journal}
  {Physical Review Materials}\ }\textbf {\bibinfo {volume} {7}},\ \bibinfo
  {pages} {016201} (\bibinfo {year} {2023})}\BibitemShut {NoStop}%
\bibitem [{\citenamefont {Plissard}\ \emph {et~al.}(2013)\citenamefont
  {Plissard}, \citenamefont {Van~Weperen}, \citenamefont {Car}, \citenamefont
  {Verheijen}, \citenamefont {Immink}, \citenamefont {Kammhuber}, \citenamefont
  {Cornelissen}, \citenamefont {Szombati}, \citenamefont {Geresdi},
  \citenamefont {Frolov} \emph {et~al.}}]{plissard2013formation}%
  \BibitemOpen
  \bibfield  {author} {\bibinfo {author} {\bibfnamefont {S.~R.}\ \bibnamefont
  {Plissard}}, \bibinfo {author} {\bibfnamefont {I.}~\bibnamefont
  {Van~Weperen}}, \bibinfo {author} {\bibfnamefont {D.}~\bibnamefont {Car}},
  \bibinfo {author} {\bibfnamefont {M.~A.}\ \bibnamefont {Verheijen}}, \bibinfo
  {author} {\bibfnamefont {G.~W.}\ \bibnamefont {Immink}}, \bibinfo {author}
  {\bibfnamefont {J.}~\bibnamefont {Kammhuber}}, \bibinfo {author}
  {\bibfnamefont {L.~J.}\ \bibnamefont {Cornelissen}}, \bibinfo {author}
  {\bibfnamefont {D.~B.}\ \bibnamefont {Szombati}}, \bibinfo {author}
  {\bibfnamefont {A.}~\bibnamefont {Geresdi}}, \bibinfo {author} {\bibfnamefont
  {S.~M.}\ \bibnamefont {Frolov}},  \emph {et~al.},\ }\href@noop {} {\bibfield
  {journal} {\bibinfo  {journal} {Nature Nanotechnology}\ }\textbf {\bibinfo
  {volume} {8}},\ \bibinfo {pages} {859} (\bibinfo {year} {2013})}\BibitemShut
  {NoStop}%
\bibitem [{\citenamefont {Stampfer}\ \emph {et~al.}(2022)\citenamefont
  {Stampfer}, \citenamefont {Carrad}, \citenamefont {Olsteins}, \citenamefont
  {Petersen}, \citenamefont {Khan}, \citenamefont {Krogstrup},\ and\
  \citenamefont {Jespersen}}]{stampfer2022andreev}%
  \BibitemOpen
  \bibfield  {author} {\bibinfo {author} {\bibfnamefont {L.}~\bibnamefont
  {Stampfer}}, \bibinfo {author} {\bibfnamefont {D.~J.}\ \bibnamefont
  {Carrad}}, \bibinfo {author} {\bibfnamefont {D.}~\bibnamefont {Olsteins}},
  \bibinfo {author} {\bibfnamefont {C.~E.}\ \bibnamefont {Petersen}}, \bibinfo
  {author} {\bibfnamefont {S.~A.}\ \bibnamefont {Khan}}, \bibinfo {author}
  {\bibfnamefont {P.}~\bibnamefont {Krogstrup}}, \ and\ \bibinfo {author}
  {\bibfnamefont {T.~S.}\ \bibnamefont {Jespersen}},\ }\href@noop {} {\bibfield
   {journal} {\bibinfo  {journal} {Advanced Materials}\ }\textbf {\bibinfo
  {volume} {34}},\ \bibinfo {pages} {2108878} (\bibinfo {year}
  {2022})}\BibitemShut {NoStop}%
\bibitem [{\citenamefont {Wang}\ \emph {et~al.}(2002)\citenamefont {Wang},
  \citenamefont {Wang}, \citenamefont {Brown}, \citenamefont {Brown},\ and\
  \citenamefont {May}}]{wang2002thermodynamic}%
  \BibitemOpen
  \bibfield  {author} {\bibinfo {author} {\bibfnamefont {Y.}~\bibnamefont
  {Wang}}, \bibinfo {author} {\bibfnamefont {Z.}~\bibnamefont {Wang}}, \bibinfo
  {author} {\bibfnamefont {T.}~\bibnamefont {Brown}}, \bibinfo {author}
  {\bibfnamefont {A.}~\bibnamefont {Brown}}, \ and\ \bibinfo {author}
  {\bibfnamefont {G.}~\bibnamefont {May}},\ }\href@noop {} {\bibfield
  {journal} {\bibinfo  {journal} {Journal of crystal growth}\ }\textbf
  {\bibinfo {volume} {242}},\ \bibinfo {pages} {5} (\bibinfo {year}
  {2002})}\BibitemShut {NoStop}%
\bibitem [{\citenamefont {Kaspi}(1999)}]{kaspi1999compositional}%
  \BibitemOpen
  \bibfield  {author} {\bibinfo {author} {\bibfnamefont {R.}~\bibnamefont
  {Kaspi}},\ }\href@noop {} {\bibfield  {journal} {\bibinfo  {journal} {Journal
  of crystal growth}\ }\textbf {\bibinfo {volume} {201}},\ \bibinfo {pages}
  {864} (\bibinfo {year} {1999})}\BibitemShut {NoStop}%
\bibitem [{\citenamefont {Xie}\ \emph {et~al.}(1999)\citenamefont {Xie},
  \citenamefont {Van~Nostrand}, \citenamefont {Brown},\ and\ \citenamefont
  {Stutz}}]{xie1999arsenic}%
  \BibitemOpen
  \bibfield  {author} {\bibinfo {author} {\bibfnamefont {Q.}~\bibnamefont
  {Xie}}, \bibinfo {author} {\bibfnamefont {J.}~\bibnamefont {Van~Nostrand}},
  \bibinfo {author} {\bibfnamefont {J.}~\bibnamefont {Brown}}, \ and\ \bibinfo
  {author} {\bibfnamefont {C.}~\bibnamefont {Stutz}},\ }\href@noop {}
  {\bibfield  {journal} {\bibinfo  {journal} {Journal of Applied Physics}\
  }\textbf {\bibinfo {volume} {86}},\ \bibinfo {pages} {329} (\bibinfo {year}
  {1999})}\BibitemShut {NoStop}%
\bibitem [{\citenamefont {Anderson}\ and\ \citenamefont
  {Millunchick}(2018)}]{anderson2018atomistic}%
  \BibitemOpen
  \bibfield  {author} {\bibinfo {author} {\bibfnamefont {E.~M.}\ \bibnamefont
  {Anderson}}\ and\ \bibinfo {author} {\bibfnamefont {J.~M.}\ \bibnamefont
  {Millunchick}},\ }\href@noop {} {\bibfield  {journal} {\bibinfo  {journal}
  {Surface Science}\ }\textbf {\bibinfo {volume} {667}},\ \bibinfo {pages} {45}
  (\bibinfo {year} {2018})}\BibitemShut {NoStop}%
\bibitem [{\citenamefont {Draelos}\ \emph
  {et~al.}(2019{\natexlab{b}})\citenamefont {Draelos}, \citenamefont {Wei},
  \citenamefont {Seredinski}, \citenamefont {Li}, \citenamefont {Mehta},
  \citenamefont {Watanabe}, \citenamefont {Taniguchi}, \citenamefont
  {Borzenets}, \citenamefont {Amet},\ and\ \citenamefont
  {Finkelstein}}]{Draelos:2019}%
  \BibitemOpen
  \bibfield  {author} {\bibinfo {author} {\bibfnamefont {A.~W.}\ \bibnamefont
  {Draelos}}, \bibinfo {author} {\bibfnamefont {M.-T.}\ \bibnamefont {Wei}},
  \bibinfo {author} {\bibfnamefont {A.}~\bibnamefont {Seredinski}}, \bibinfo
  {author} {\bibfnamefont {H.}~\bibnamefont {Li}}, \bibinfo {author}
  {\bibfnamefont {Y.}~\bibnamefont {Mehta}}, \bibinfo {author} {\bibfnamefont
  {K.}~\bibnamefont {Watanabe}}, \bibinfo {author} {\bibfnamefont
  {T.}~\bibnamefont {Taniguchi}}, \bibinfo {author} {\bibfnamefont {I.~V.}\
  \bibnamefont {Borzenets}}, \bibinfo {author} {\bibfnamefont {F.}~\bibnamefont
  {Amet}}, \ and\ \bibinfo {author} {\bibfnamefont {G.}~\bibnamefont
  {Finkelstein}},\ }\href {\doibase 10.1021/acs.nanolett.8b04330} {\bibfield
  {journal} {\bibinfo  {journal} {Nano Lett.}\ }\textbf {\bibinfo {volume}
  {19}},\ \bibinfo {pages} {1039} (\bibinfo {year}
  {2019}{\natexlab{b}})}\BibitemShut {NoStop}%
\bibitem [{\citenamefont {Jaklevic}\ \emph {et~al.}(1964)\citenamefont
  {Jaklevic}, \citenamefont {Lambe}, \citenamefont {Silver},\ and\
  \citenamefont {Mercereau}}]{Jaklevic1964Quantum}%
  \BibitemOpen
  \bibfield  {author} {\bibinfo {author} {\bibfnamefont {R.}~\bibnamefont
  {Jaklevic}}, \bibinfo {author} {\bibfnamefont {J.}~\bibnamefont {Lambe}},
  \bibinfo {author} {\bibfnamefont {A.}~\bibnamefont {Silver}}, \ and\ \bibinfo
  {author} {\bibfnamefont {J.}~\bibnamefont {Mercereau}},\ }\href@noop {}
  {\bibfield  {journal} {\bibinfo  {journal} {Physical Review Letters}\
  }\textbf {\bibinfo {volume} {12}},\ \bibinfo {pages} {158} (\bibinfo {year}
  {1964})}\BibitemShut {NoStop}%
\bibitem [{\citenamefont {Giazotto}\ \emph {et~al.}(2011)\citenamefont
  {Giazotto}, \citenamefont {Spathis}, \citenamefont {Roddaro}, \citenamefont
  {Biswas}, \citenamefont {Taddei}, \citenamefont {Governale},\ and\
  \citenamefont {Sorba}}]{giazotto2011josephson}%
  \BibitemOpen
  \bibfield  {author} {\bibinfo {author} {\bibfnamefont {F.}~\bibnamefont
  {Giazotto}}, \bibinfo {author} {\bibfnamefont {P.}~\bibnamefont {Spathis}},
  \bibinfo {author} {\bibfnamefont {S.}~\bibnamefont {Roddaro}}, \bibinfo
  {author} {\bibfnamefont {S.}~\bibnamefont {Biswas}}, \bibinfo {author}
  {\bibfnamefont {F.}~\bibnamefont {Taddei}}, \bibinfo {author} {\bibfnamefont
  {M.}~\bibnamefont {Governale}}, \ and\ \bibinfo {author} {\bibfnamefont
  {L.}~\bibnamefont {Sorba}},\ }\href@noop {} {\bibfield  {journal} {\bibinfo
  {journal} {Nature Physics}\ }\textbf {\bibinfo {volume} {7}},\ \bibinfo
  {pages} {857} (\bibinfo {year} {2011})}\BibitemShut {NoStop}%
\bibitem [{\citenamefont {Savinov}(2016)}]{savinov2016enhancement}%
  \BibitemOpen
  \bibfield  {author} {\bibinfo {author} {\bibfnamefont {D.}~\bibnamefont
  {Savinov}},\ }\href@noop {} {\bibfield  {journal} {\bibinfo  {journal}
  {Physica C: Superconductivity and its Applications}\ }\textbf {\bibinfo
  {volume} {527}},\ \bibinfo {pages} {80} (\bibinfo {year} {2016})}\BibitemShut
  {NoStop}%
\bibitem [{\citenamefont {Galaktionov}\ and\ \citenamefont
  {Zaikin}(2013)}]{galaktionov2013current}%
  \BibitemOpen
  \bibfield  {author} {\bibinfo {author} {\bibfnamefont {A.~V.}\ \bibnamefont
  {Galaktionov}}\ and\ \bibinfo {author} {\bibfnamefont {A.~D.}\ \bibnamefont
  {Zaikin}},\ }\href@noop {} {\bibfield  {journal} {\bibinfo  {journal}
  {Physical Review B—Condensed Matter and Materials Physics}\ }\textbf
  {\bibinfo {volume} {88}},\ \bibinfo {pages} {104513} (\bibinfo {year}
  {2013})}\BibitemShut {NoStop}%
\bibitem [{\citenamefont {Doh}\ \emph {et~al.}(2005)\citenamefont {Doh},
  \citenamefont {Dam}, \citenamefont {Roest}, \citenamefont {Bakkers},
  \citenamefont {Kouwenhoven},\ and\ \citenamefont {Franceschi1}}]{Doh2005}%
  \BibitemOpen
  \bibfield  {author} {\bibinfo {author} {\bibfnamefont {Y.~J.}\ \bibnamefont
  {Doh}}, \bibinfo {author} {\bibfnamefont {J.~A.~V.}\ \bibnamefont {Dam}},
  \bibinfo {author} {\bibfnamefont {A.~L.}\ \bibnamefont {Roest}}, \bibinfo
  {author} {\bibfnamefont {E.~P. A.~M.}\ \bibnamefont {Bakkers}}, \bibinfo
  {author} {\bibfnamefont {L.~P.}\ \bibnamefont {Kouwenhoven}}, \ and\ \bibinfo
  {author} {\bibfnamefont {S.~D.}\ \bibnamefont {Franceschi1}},\ }\href@noop {}
  {\bibfield  {journal} {\bibinfo  {journal} {Science}\ }\textbf {\bibinfo
  {volume} {309}},\ \bibinfo {pages} {272} (\bibinfo {year}
  {2005})}\BibitemShut {NoStop}%
\bibitem [{\citenamefont {Spanton}\ \emph {et~al.}(2017)\citenamefont
  {Spanton}, \citenamefont {Deng}, \citenamefont {Vaitiek{\.e}nas},
  \citenamefont {Krogstrup}, \citenamefont {Nyg{\aa}rd}, \citenamefont
  {Marcus},\ and\ \citenamefont {Moler}}]{spanton2017current}%
  \BibitemOpen
  \bibfield  {author} {\bibinfo {author} {\bibfnamefont {E.~M.}\ \bibnamefont
  {Spanton}}, \bibinfo {author} {\bibfnamefont {M.}~\bibnamefont {Deng}},
  \bibinfo {author} {\bibfnamefont {S.}~\bibnamefont {Vaitiek{\.e}nas}},
  \bibinfo {author} {\bibfnamefont {P.}~\bibnamefont {Krogstrup}}, \bibinfo
  {author} {\bibfnamefont {J.}~\bibnamefont {Nyg{\aa}rd}}, \bibinfo {author}
  {\bibfnamefont {C.~M.}\ \bibnamefont {Marcus}}, \ and\ \bibinfo {author}
  {\bibfnamefont {K.~A.}\ \bibnamefont {Moler}},\ }\href@noop {} {\bibfield
  {journal} {\bibinfo  {journal} {Nature Physics}\ }\textbf {\bibinfo {volume}
  {13}},\ \bibinfo {pages} {1177} (\bibinfo {year} {2017})}\BibitemShut
  {NoStop}%
\bibitem [{\citenamefont {Sestoft}\ \emph {et~al.}(2018)\citenamefont
  {Sestoft}, \citenamefont {Kanne}, \citenamefont {Gejl}, \citenamefont {{von
  Soosten}}, \citenamefont {Yodh}, \citenamefont {Sherman}, \citenamefont
  {Tarasinski}, \citenamefont {Wimmer}, \citenamefont {Johnson}, \citenamefont
  {Deng}, \citenamefont {Nyg{\aa}rd}, \citenamefont {Jespersen}, \citenamefont
  {Marcus},\ and\ \citenamefont {Krogstrup}}]{sestoft:2018}%
  \BibitemOpen
  \bibfield  {author} {\bibinfo {author} {\bibfnamefont {J.~E.}\ \bibnamefont
  {Sestoft}}, \bibinfo {author} {\bibfnamefont {T.}~\bibnamefont {Kanne}},
  \bibinfo {author} {\bibfnamefont {A.~N.}\ \bibnamefont {Gejl}}, \bibinfo
  {author} {\bibfnamefont {M.}~\bibnamefont {{von Soosten}}}, \bibinfo {author}
  {\bibfnamefont {J.~S.}\ \bibnamefont {Yodh}}, \bibinfo {author}
  {\bibfnamefont {D.}~\bibnamefont {Sherman}}, \bibinfo {author} {\bibfnamefont
  {B.}~\bibnamefont {Tarasinski}}, \bibinfo {author} {\bibfnamefont
  {M.}~\bibnamefont {Wimmer}}, \bibinfo {author} {\bibfnamefont
  {E.}~\bibnamefont {Johnson}}, \bibinfo {author} {\bibfnamefont
  {M.}~\bibnamefont {Deng}}, \bibinfo {author} {\bibfnamefont {J.}~\bibnamefont
  {Nyg{\aa}rd}}, \bibinfo {author} {\bibfnamefont {T.~S.}\ \bibnamefont
  {Jespersen}}, \bibinfo {author} {\bibfnamefont {C.~M.}\ \bibnamefont
  {Marcus}}, \ and\ \bibinfo {author} {\bibfnamefont {P.}~\bibnamefont
  {Krogstrup}},\ }\href {\doibase 10.1103/PhysRevMaterials.2.044202} {\bibfield
   {journal} {\bibinfo  {journal} {Phys. Rev. Materials}\ }\textbf {\bibinfo
  {volume} {2}},\ \bibinfo {pages} {044202} (\bibinfo {year}
  {2018})}\BibitemShut {NoStop}%
\bibitem [{\citenamefont {Ciancio}\ \emph {et~al.}(2022)\citenamefont
  {Ciancio}, \citenamefont {Dunin-Borkowski}, \citenamefont {Snoeck},
  \citenamefont {Kociak}, \citenamefont {Holmestad}, \citenamefont {Verbeeck},
  \citenamefont {Kirkland}, \citenamefont {Kothleitner},\ and\ \citenamefont
  {Arbiol}}]{ciancio2022dream}%
  \BibitemOpen
  \bibfield  {author} {\bibinfo {author} {\bibfnamefont {R.}~\bibnamefont
  {Ciancio}}, \bibinfo {author} {\bibfnamefont {R.~E.}\ \bibnamefont
  {Dunin-Borkowski}}, \bibinfo {author} {\bibfnamefont {E.}~\bibnamefont
  {Snoeck}}, \bibinfo {author} {\bibfnamefont {M.}~\bibnamefont {Kociak}},
  \bibinfo {author} {\bibfnamefont {R.}~\bibnamefont {Holmestad}}, \bibinfo
  {author} {\bibfnamefont {J.}~\bibnamefont {Verbeeck}}, \bibinfo {author}
  {\bibfnamefont {A.~I.}\ \bibnamefont {Kirkland}}, \bibinfo {author}
  {\bibfnamefont {G.}~\bibnamefont {Kothleitner}}, \ and\ \bibinfo {author}
  {\bibfnamefont {J.}~\bibnamefont {Arbiol}},\ }\href@noop {} {\bibfield
  {journal} {\bibinfo  {journal} {Microscopy and Microanalysis}\ }\textbf
  {\bibinfo {volume} {28}},\ \bibinfo {pages} {2900} (\bibinfo {year}
  {2022})}\BibitemShut {NoStop}%
\end{thebibliography}%

\end{document}


\newpage

\section{S1 - Energy dispersive x-ray spectroscopy (EDX) to analyze the NC composition}

\begin{figure*}[ht!]
    \centering
    \includegraphics[scale=0.7]{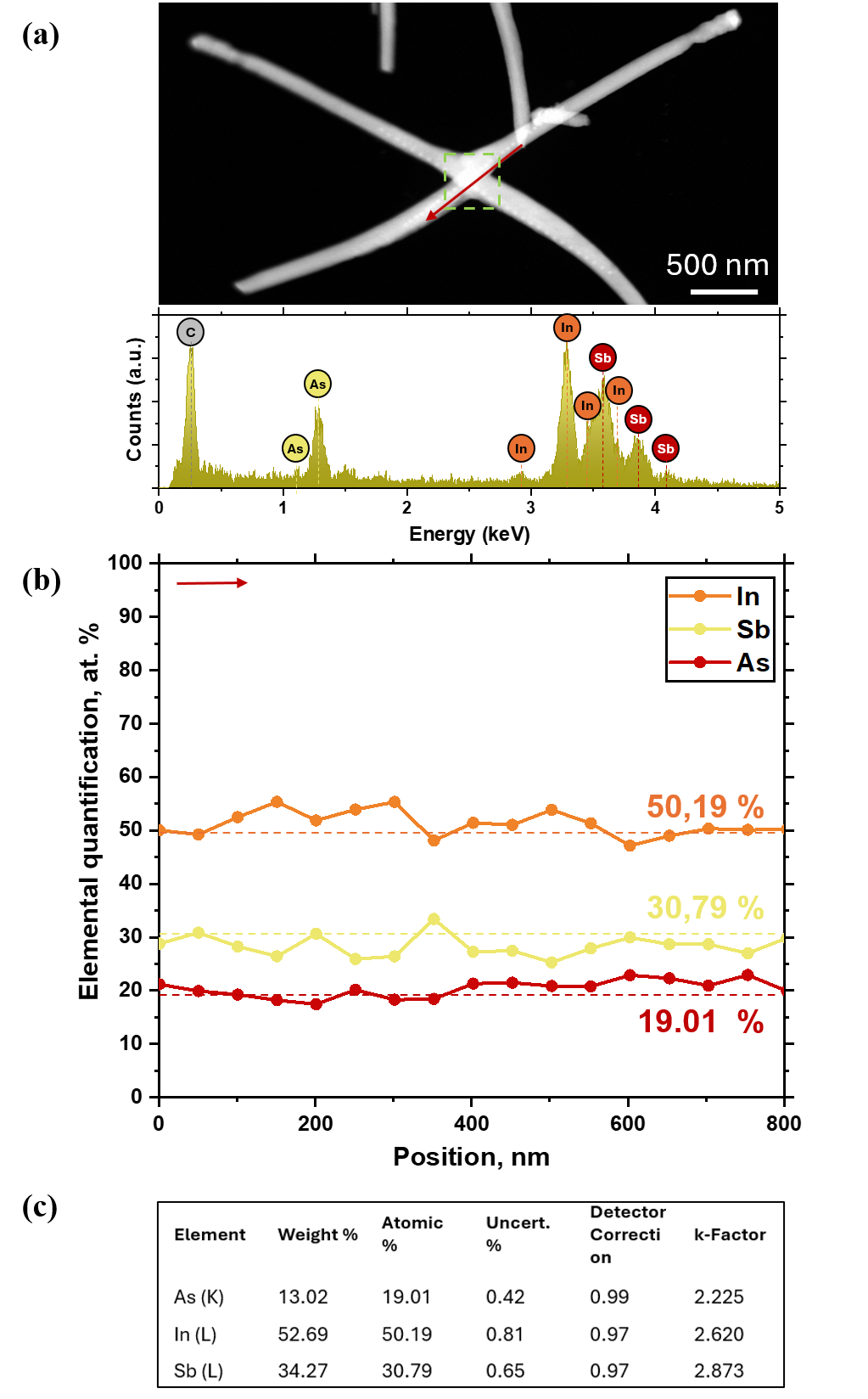}
    \caption{(a), SEM image of the analyzed NC for EDX and the elements observed in the green boxed area is presented in the energy scale below. (b-c), Quantitative profile of the elements present in the NC and the composition is- In: 50.19 $\%$, Sb: 30.79 and As 19.01 $\%$.}
    \label{SI_fig1_EDX}
\end{figure*}

\newpage

\section{S2 - Additional Analysis in the Edge of the junction area}

\begin{figure*}[ht!]
    \centering
    \includegraphics[scale=0.6]{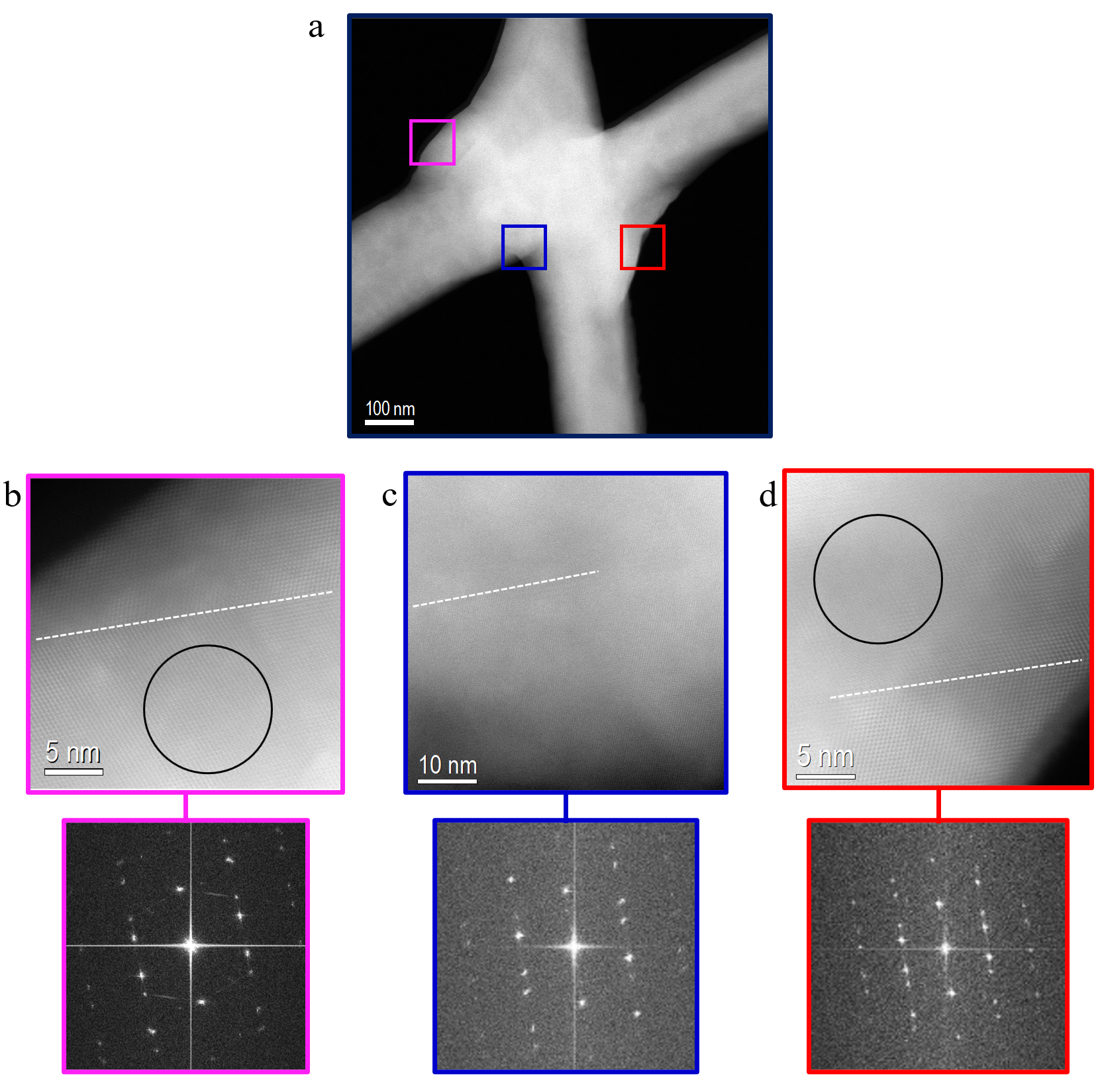}
    \caption{(a), Zoomed out Transmission electron microscopy (TEM) image of the junction area. (b-d), The bottom panels correspond to HAADF micrographs obtained at the edges of the junction where twins in the (-1-1-1) plane can be observed . The analyzed areas are shown in different colours in panel (a). In the circled areas, the atomic positions look to be blurred. This is likely arising from the superposition of the different crystal orientations in depth, so probably there are some extra boundaries that we are not visualizing (out of axis).}
    \label{SI_fig2_Edge}
\end{figure*}

\newpage

\section{S3 - Additional Analysis in the Overgrown region of the junction area}

\begin{figure*}[ht!]
    \centering
    \includegraphics[scale=0.4]{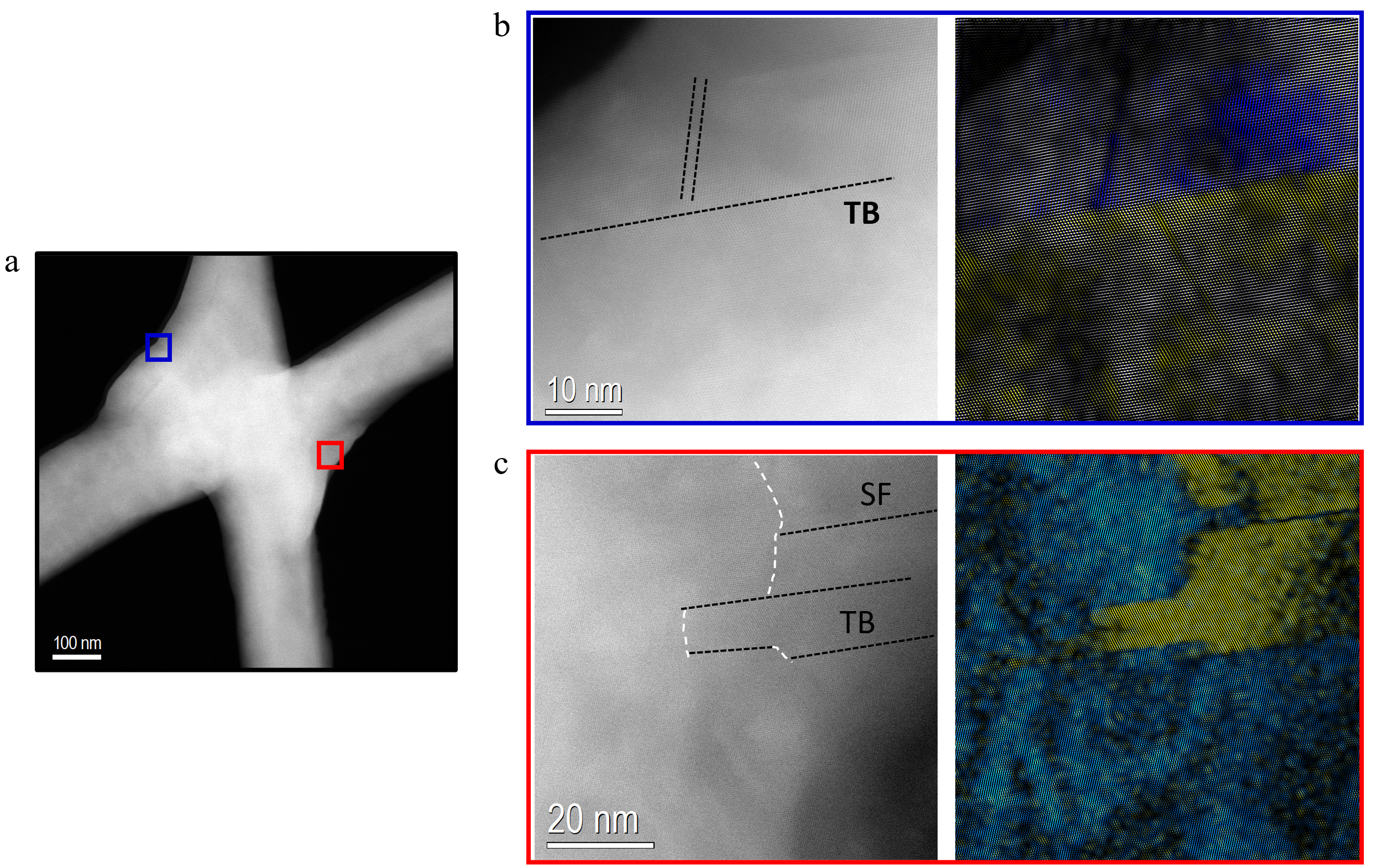}
    \caption{(a), TEM image of the junction with focusing in two overgrown edges of the junction (blue and red). (b-c), In these highlighted overgrown areas we observed more complex defective structure. In panel (b), we observed a single twinning in the (-1-11) which is also clearly visible in Fourier filtered structural map of the analyzed area. In panel (c), we also observed twin boundary, additional stacking faults and some irregular boundaries that arises from different orientational merging. It was not possible to resolve the dumbbell pairs due to the the strong thickness variations, but also likely that these irregular boundaries have some polar shift, given the shift in [11-2] direction produced by the twin boundaries.}
    \label{SI_fig3_EO}
\end{figure*}

\newpage

\section{S4 - Compression in the overgrown area of the Nanocross Junction}

\begin{figure*}[ht!]
    \centering
    \includegraphics[scale=0.6]{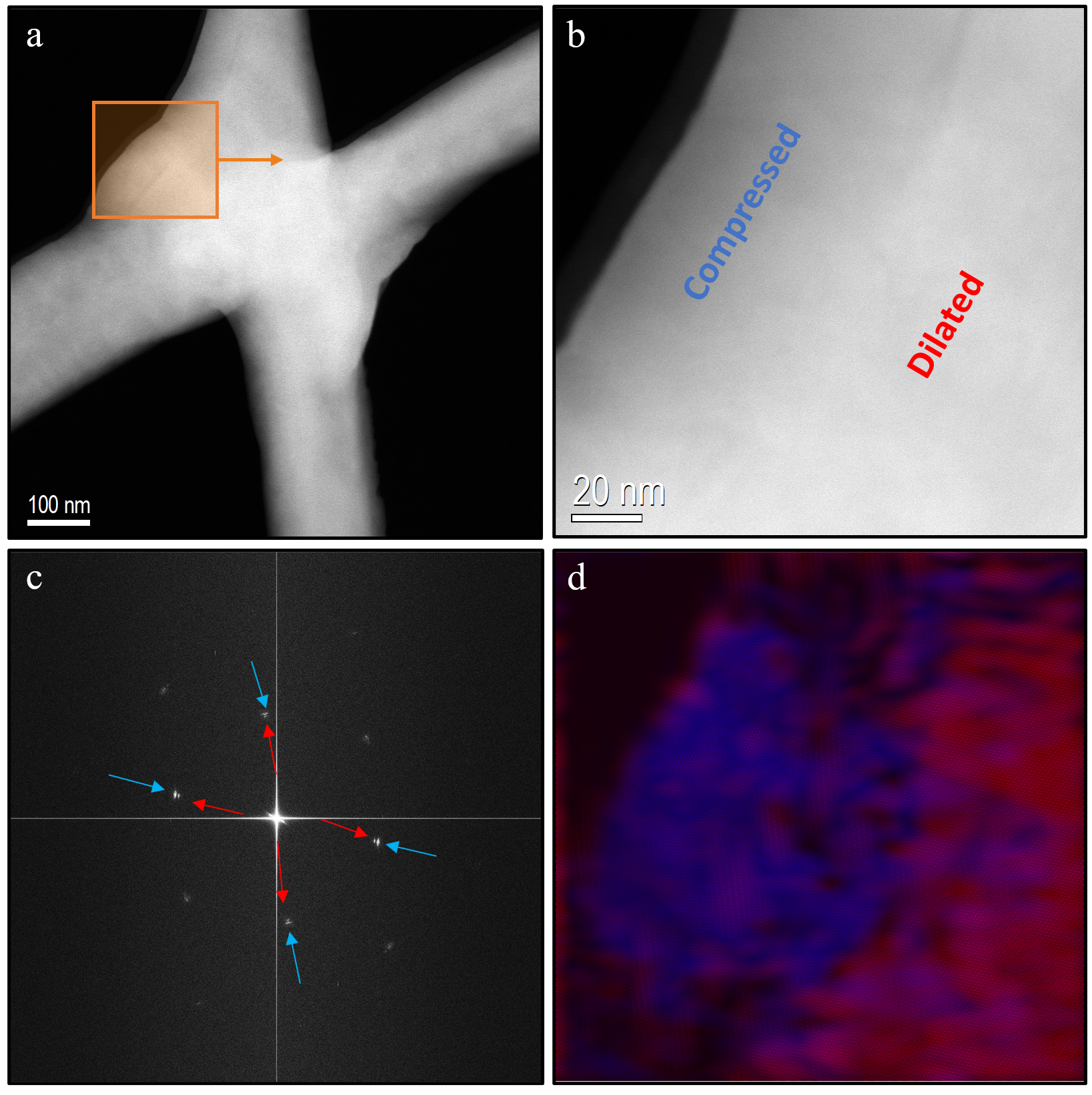}
    \caption{(a), TEM image of the junction with the focus in the overgrown area. (b-c), Zoomed in TEM image and the diffraction of the focused overgrown area. The outer region of the NC within the overgrown area exhibits a compressive strain relative to the central part of the cross, which may be related to compositional fluctuations, although such variations were not detected in the EELS maps. (d), FFT analysis of the overall compression is performed in the focused overgrowth area.}
    \label{SI_fig4_Alcom}
\end{figure*}

\newpage

\section{S5 - Additional data on other devices}

\begin{figure*}[ht!]
    \centering
    \includegraphics{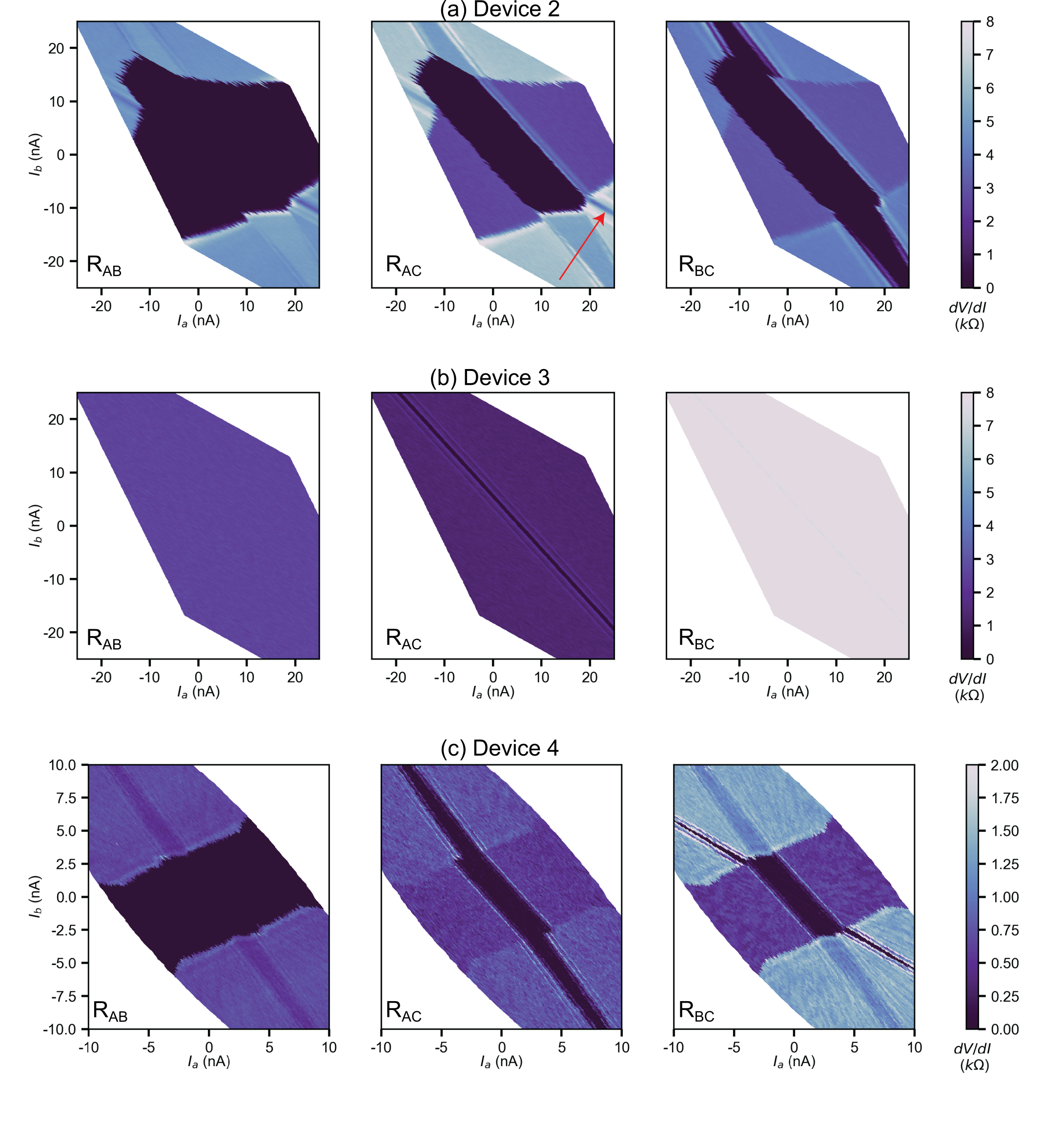}
    \caption{Differential resistance $R_\mathrm{AB}$, $R_\mathrm{AC}$, and $R_\mathrm{BC}$ for three other devices.}
    \label{fig:SUP4_otherdevices}
\end{figure*}

Fig. \ref{fig:SUP4_otherdevices} shows the differential conductance data of three additional devices from the same chip. Device 4 shows, analogous to the device presented in the main text fluxdependence of the center region as well as of $\frac{dV_{\mathrm{AC}}}{dI_{\mathrm{AC}}}$ and $\frac{dV_{\mathrm{BC}}}{dI_{\mathrm{BC}}}$, while Device 2 show negligible dependence of the critical current contours on the flux, Device 3 does not show all three superconducting connections.

\section*{S6 - Differential resistance maps with different sweep directions}

\begin{figure*}[ht!]
    \centering
    \includegraphics[scale=0.8]{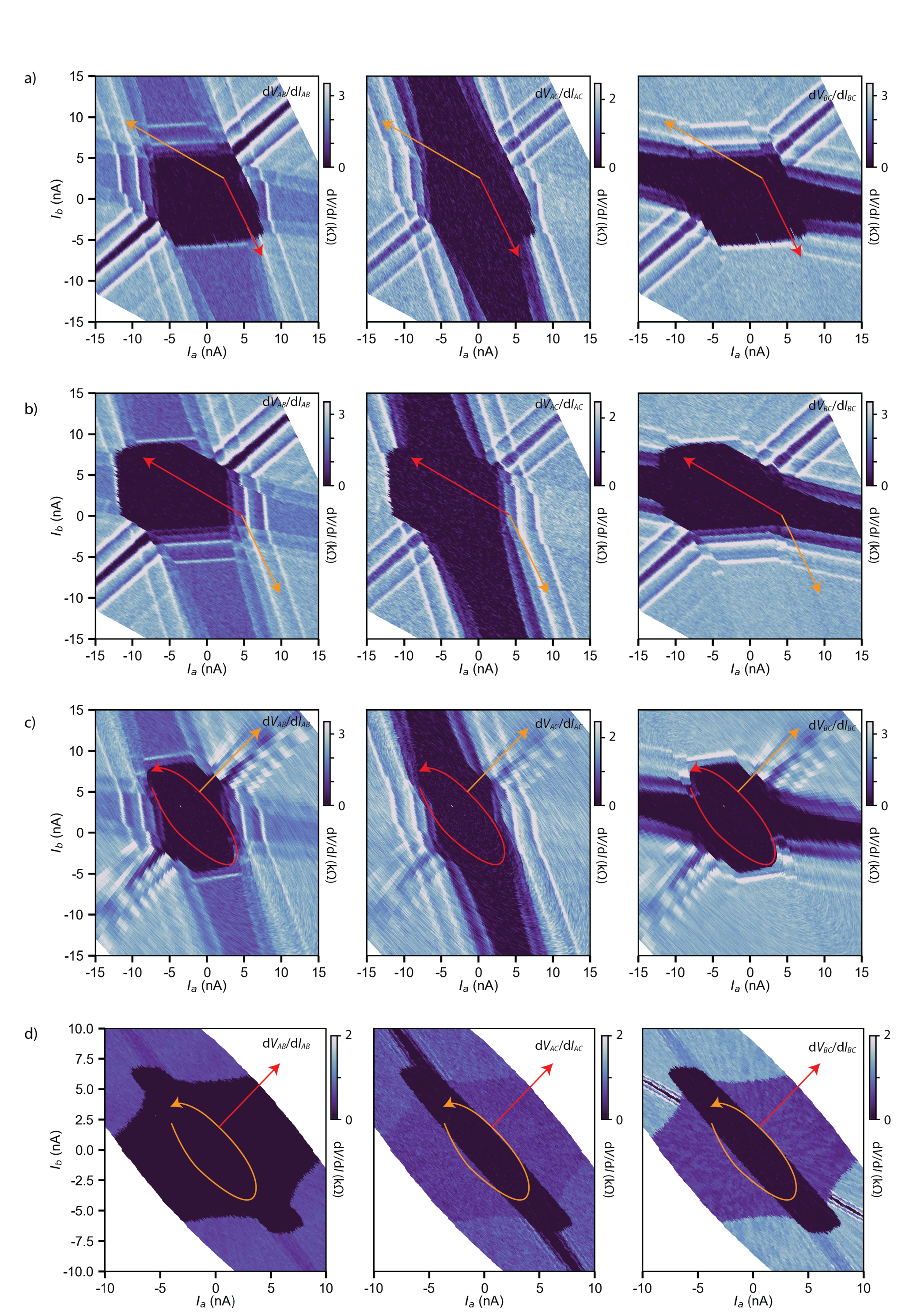}
    \caption{Influence of the sweeping direction of the switching and retrapping currents. (a) Differential resistance maps as shown in the main text. The red arrow indicates the sweep direction, the orange arrow the step direction. (b) Same as (a), but with inverted sweep and step direction. (c) Differential resistance map of an analogous device with an angular sweep and radial step direction. (d) Same as (c), but with radial sweep and angular step direction.}
    \label{fig:SUP1_sweepdirection}
\end{figure*}

In a two terminal Josephson Junction, the RCSJ model describes the dynamics of the phase with the tilted washboard model in which the switching and retrapping currents arise as direct consequences. In a Josephson junction characterized by two independent currents and phases, the washboard potential depends on two phase variables, and the corresponding dynamics are described by a trajectory in two-dimensional phase space. As a result, the switching and retrapping currents depend on the sweeping direction of the two bias currents.

We want to note that with our employed measurement setup, we are not able to apply a current at a terminal, instead we apply a voltage over a bias resistor. While those two are equivalent in a two terminal geometry, given a large enough bias resistor, in the 3 terminal case the effective current flowing through a terminal follows a linear transformation in $I_\mathrm{A}$-$(I_\mathrm{B}$ phase space, transforming squares into rhombi and circles into ellipses. It is thus not trivial to vary only one current while not altering the other. This explains the tilted and curved choice of line cuts in the main text.

In Figure \ref{fig:SUP1_sweepdirection}(a) we show the $I_\mathrm{A}$-$I_\mathrm{B}$ map sweeping the same direction as in the main text. Effectively we sweep the DC-Voltage applied at the bias resistor at terminal A and step the DC-Voltage applied at the Bias resistor at terminal B. The shape of the center region deviates from the two-fold rotation symmetry expected by the nature of a 3 terminal device.  The transition between the superconducting and normal state occurs at much higher currents than the transition between the normal and superconducting state.
Figure \ref{fig:SUP1_sweepdirection}(b) shows the same device measured with flipped sweeping and stepping voltage directions, thus the asymmetry is oriented more towards lower values of $I_a$.
Sweeping and stepping in polar coordinates allows us to cross the border of the central superconducting region to the normal region or vice versa in all directions as long as the central region is concave. A map scanned in this direction on an analogous device is shown in Figure \ref{fig:SUP1_sweepdirection}(c,d). It is easier to resolve the coexisting region of supercurrents and observe the switching current in all directions. This method has the downside that the pixel density is decreasing towards higher currents which makes it hard to resolve small superconducting rays properly.
\newpage

\section{S7 - Additional data for flux-dependence}

The main text illustrates the two extreme cases of flux in the loop with the induced phases being 0 and $\pi$. The full dataset of the three differential resistances for the different phases is depicted in Fig \ref{fig:SUP3}. 
\begin{figure*}[ht!]
  \centering
  \includegraphics[scale=0.8]{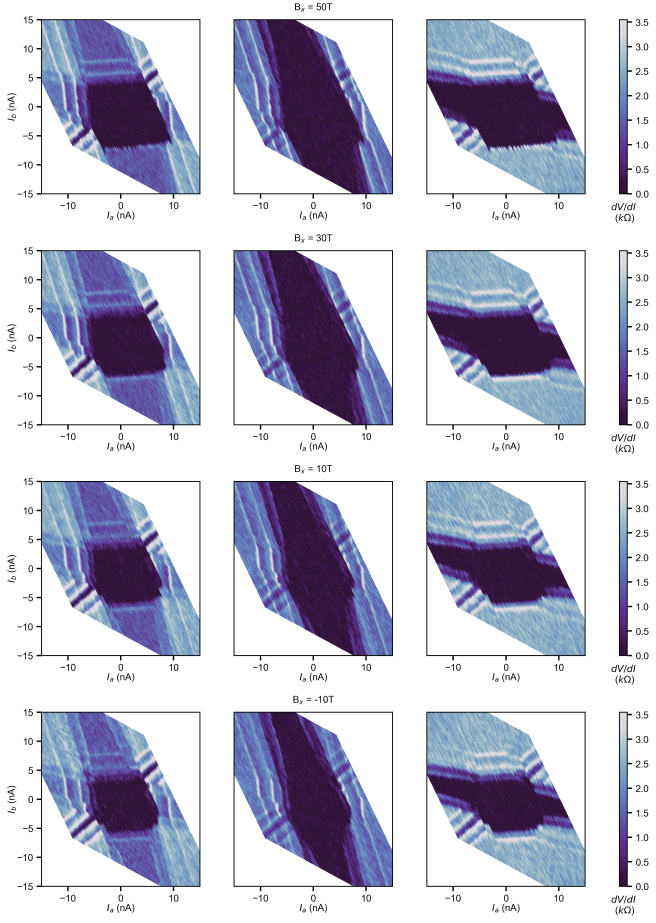}
  \caption{Differential resistances $R_\mathrm{AB}$, $R_\mathrm{AC}$, and $R_\mathrm{BC}$ as function of flux (part 1).}
  \label{fig:SUP3}
\end{figure*}

\begin{figure*}[ht!]
  \ContinuedFloat
  \centering
  \includegraphics[scale=0.8]{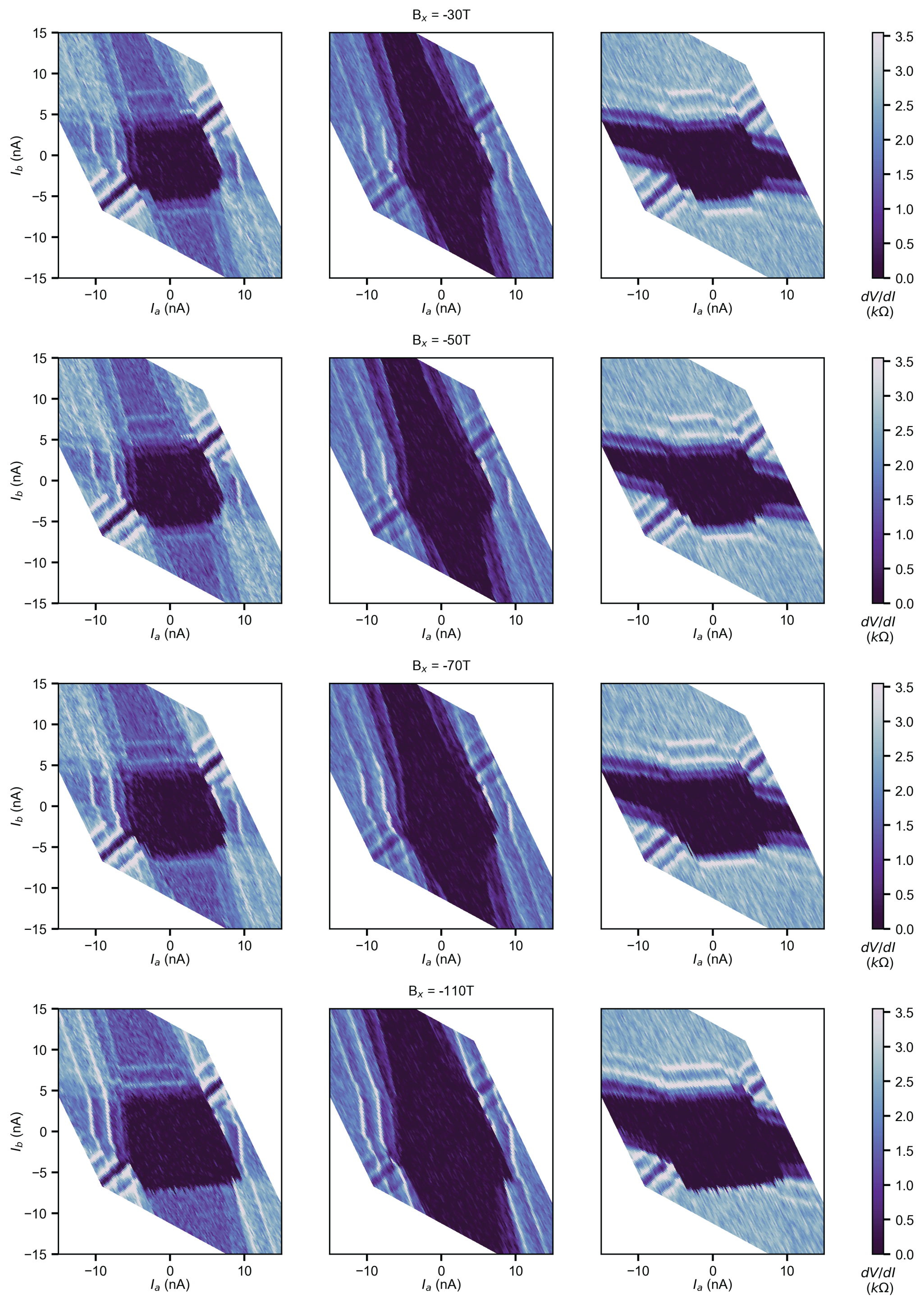}
  \caption[]{Differential resistances $R_\mathrm{AB}$, $R_\mathrm{AC}$, and $R_\mathrm{BC}$ as function of flux (part 2).}
\end{figure*}

\begin{figure*}[ht!]
  \ContinuedFloat
  \centering
  \includegraphics[scale=0.8]{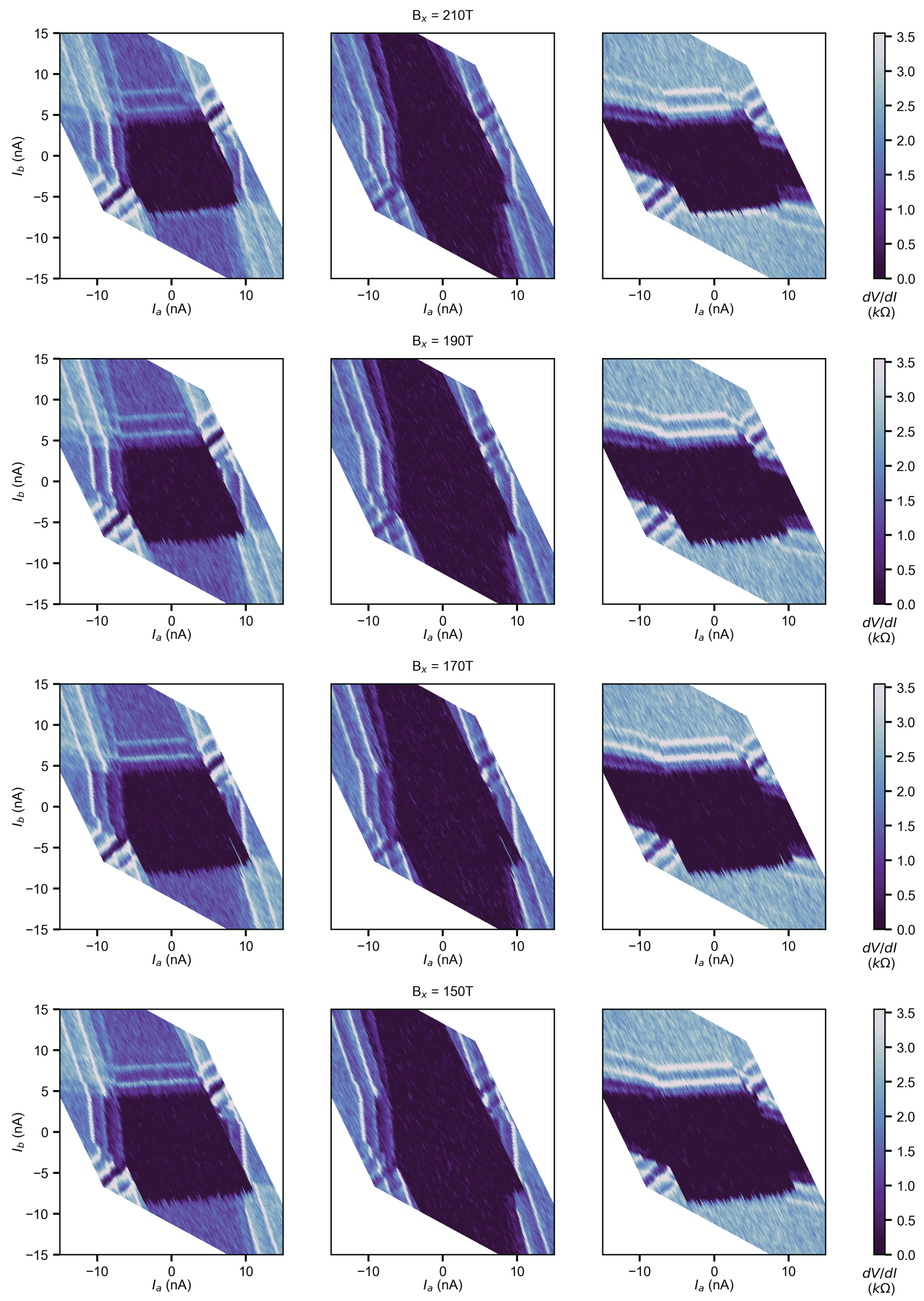}
  \caption[]{Differential resistances $R_\mathrm{AB}$, $R_\mathrm{AC}$, and $R_\mathrm{BC}$ as function of flux (part 3).}
\end{figure*}

\begin{figure*}[ht!]
  \ContinuedFloat
  \centering
  \includegraphics[scale=0.8]{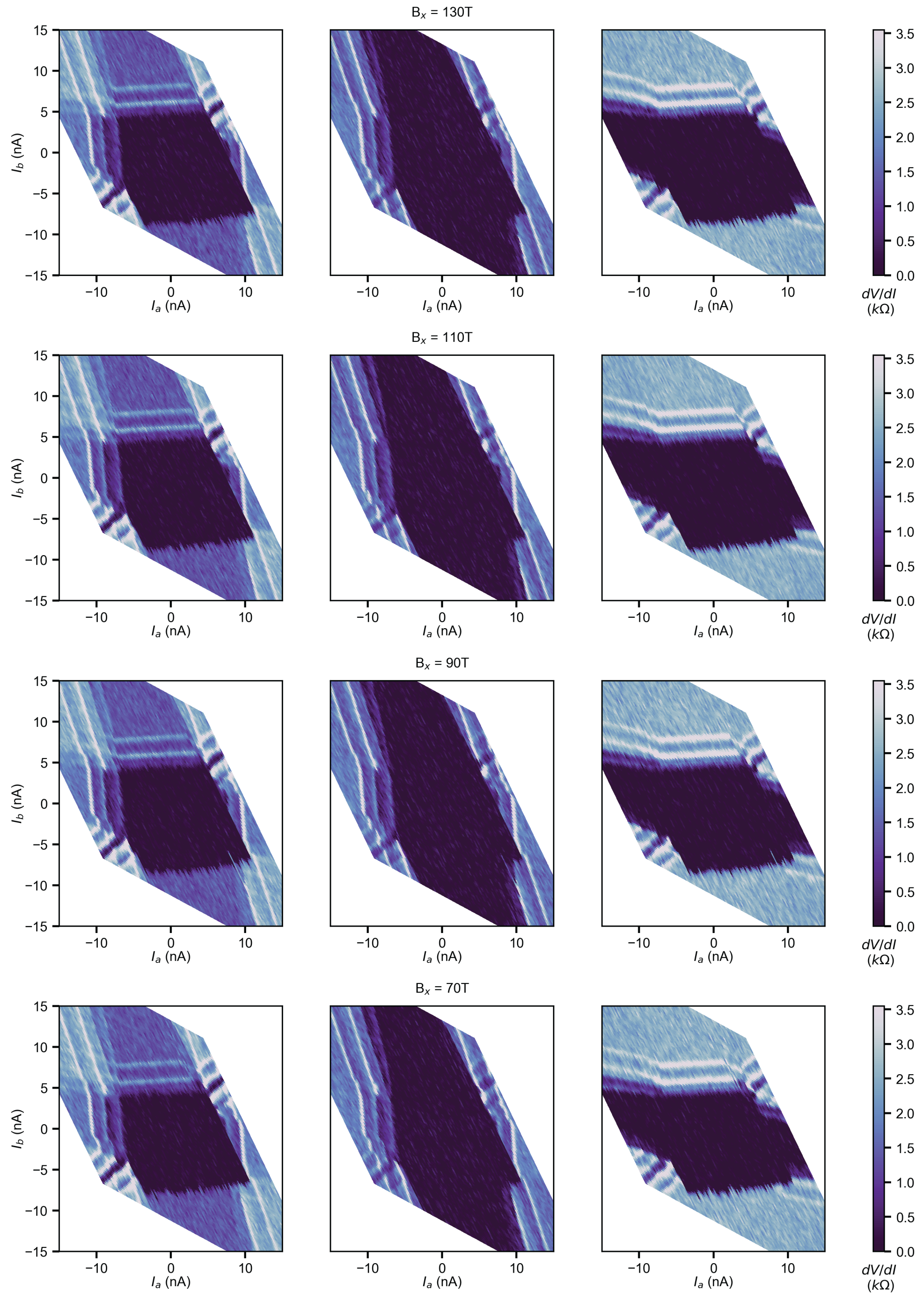}
  \caption[]{Differential resistances $R_\mathrm{AB}$, $R_\mathrm{AC}$, and $R_\mathrm{BC}$ as function of flux (part 4).}
\end{figure*}

\section{S8 - Additional data for backgate dependence}
Figures \ref{fig:SUP2_B_gatedependence} shows additional data on the differential resistances $R_\mathrm{AB}$, $R_\mathrm{AC}$, and $R_\mathrm{BC}$ for backgate voltage from 0 to 40 V and from -10 to -40 V respectively. 

\begin{figure*}[ht!]
  \centering
  \includegraphics[scale=0.8]{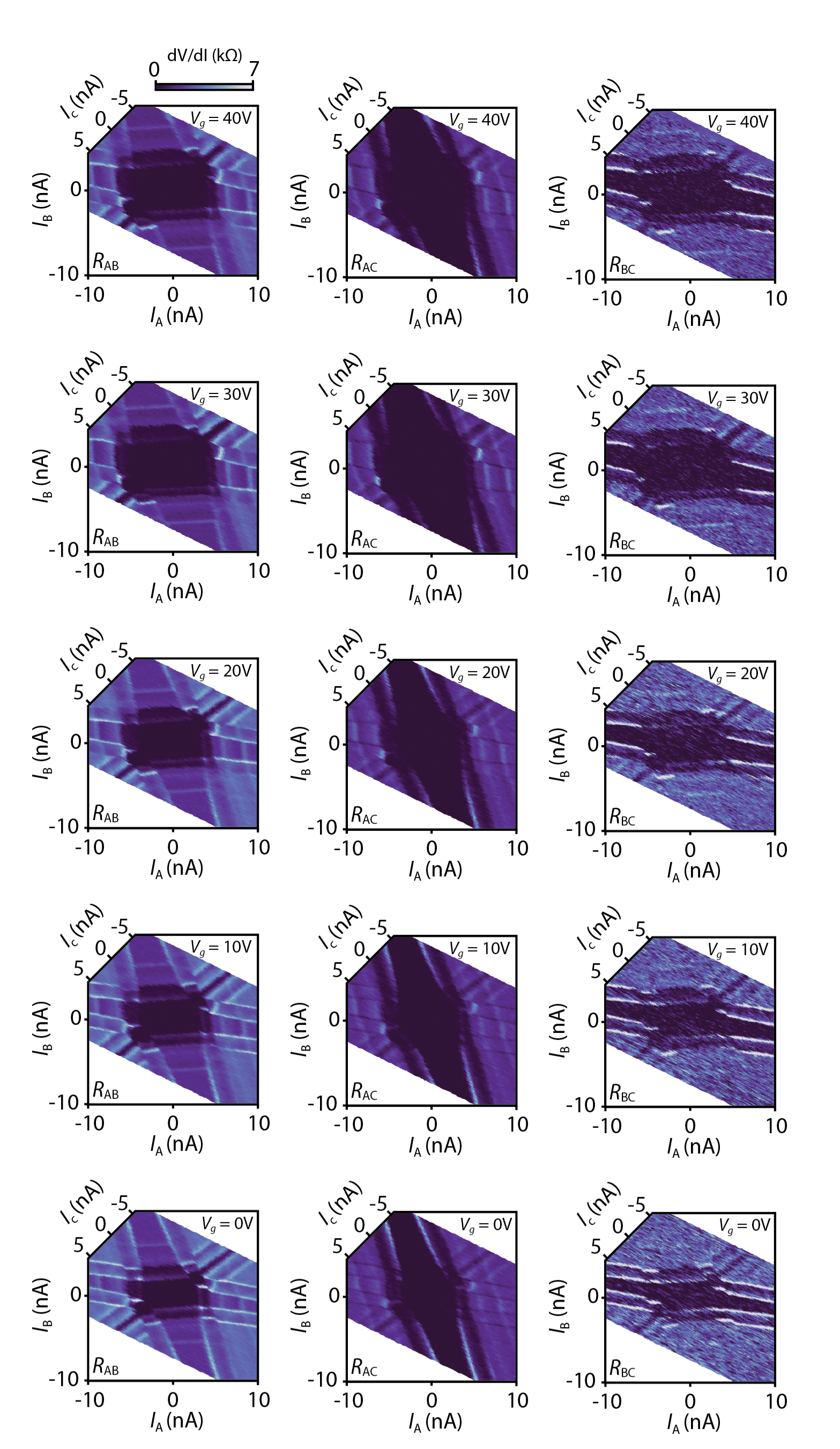}
  \caption{Differential resistances $R_\mathrm{AB}$, $R_\mathrm{AC}$, and $R_\mathrm{BC}$ for backgate voltages from -40 to 40 V. (part 1).}
  \label{fig:SUP2_B_gatedependence}
\end{figure*}

\begin{figure*}[ht!]
  \ContinuedFloat
  \centering
  \includegraphics[scale=0.8]{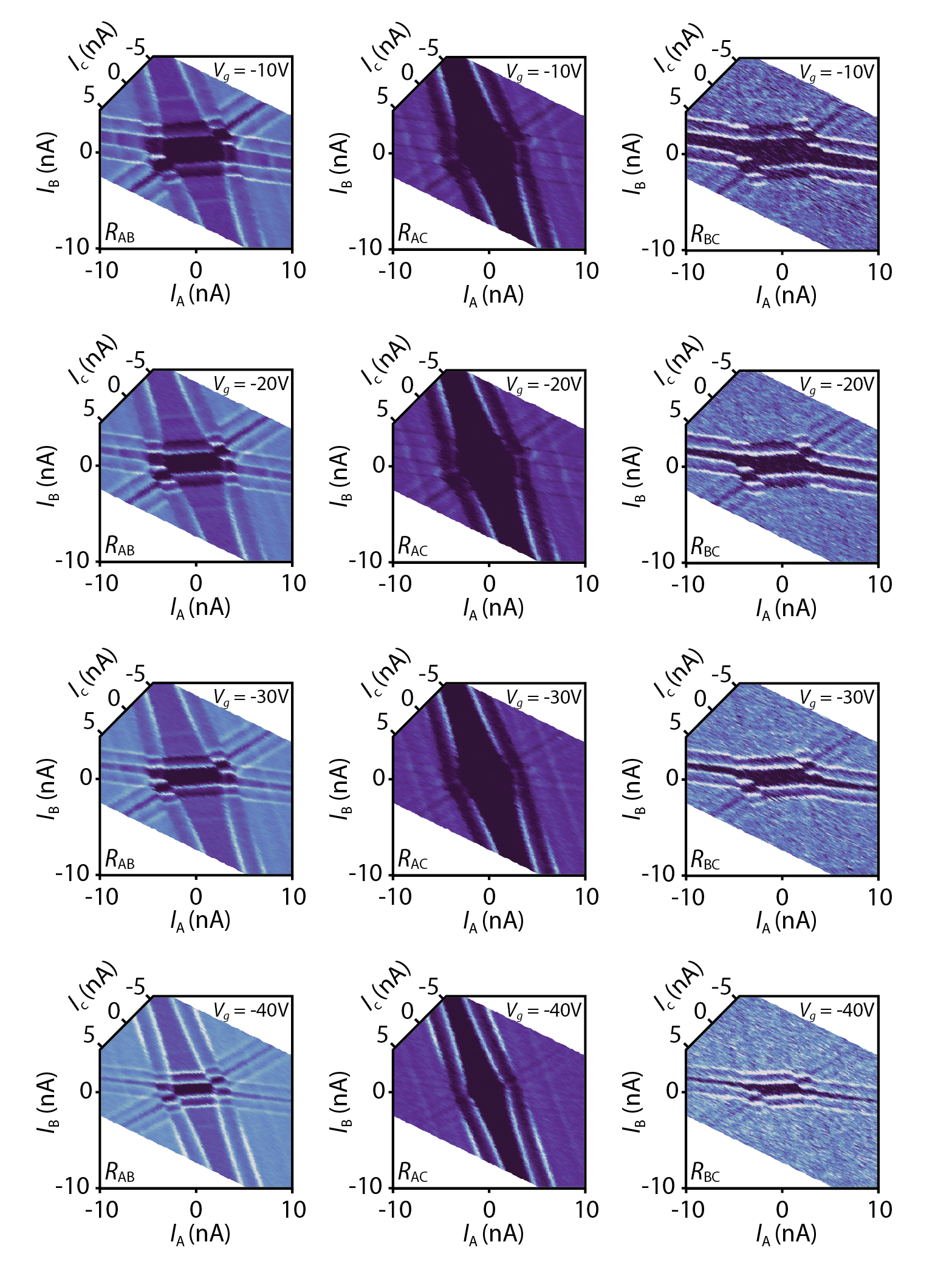}
  \caption[]{Differential resistances $R_\mathrm{AB}$, $R_\mathrm{AC}$, and $R_\mathrm{BC}$ for backgate voltages from -40 to 40 V. (part 2).}
\end{figure*}

\newpage